\documentclass[twocolumn,prl,floatfix,nofootinbib]{revtex4}

\usepackage[utf8]{inputenc}
\usepackage[T1]{fontenc} 
\usepackage[english]{babel} 

\usepackage{amsmath}
\usepackage{multirow}
\usepackage{appendix}

\usepackage{amsthm}
\usepackage{amssymb}
\usepackage[normalem]{ulem}
\usepackage{listings}
\usepackage{xcolor}
\usepackage{cancel}
\definecolor{codegreen}{rgb}{0,0.6,0}
\definecolor{codegray}{rgb}{0.5,0.5,0.5}
\definecolor{codepurple}{rgb}{0.58,0,0.82}
\definecolor{backcolour}{rgb}{0.95,0.95,0.92}

\lstdefinestyle{mystyle}{
    backgroundcolor=\color{backcolour},   
    commentstyle=\color{codegreen},
    keywordstyle=\color{magenta},
    numberstyle=\tiny\color{codegray},
    stringstyle=\color{codepurple},
    basicstyle=\ttfamily\footnotesize,
    breakatwhitespace=false,         
    breaklines=true,                 
    captionpos=b,                    
    keepspaces=true,                 
    numbersep=5pt,                  
    showspaces=false,                
    showstringspaces=false,
    showtabs=false,                  
    tabsize=1
}
\definecolor{dark-green}{RGB}{0, 128, 0}

\usepackage{enumitem}

\usepackage{dsfont}

\usepackage{graphicx}
\usepackage{color}

\usepackage{comment}
\usepackage{ stmaryrd }
\usepackage{physics}
\usepackage{tabularx}

\begin{document}

\title{$\mathbb{Z}_L$ symmetry breaking in SU(N) Fermi-Hubbard dots at zero and finite temperature.}
\author{Lo\"ic Herviou}
\affiliation{Univ. Grenoble Alpes, CNRS, LPMMC, 38000 Grenoble, France}
\author{Elodie Campan}
\affiliation{Univ. Grenoble Alpes, CNRS, LPMMC, 38000 Grenoble, France}
\author{Pierre Nataf}
\affiliation{Univ. Grenoble Alpes, CNRS, LPMMC, 38000 Grenoble, France}

\begin{abstract}
We address the SU(N) Fermi-Hubbard model on a chain, with $N$ the number of degenerate orbitals, or colors, for each fermion.
In the limit of both large number of colors $N$ and particles, and small number of sites $L \geq 2$, the model is proved to undergo a $\mathbb{Z}_L$ symmetry breaking
for attractive local interaction amplitude $U$. Using a combination of Exact Diagonalization with full SU(N) symmetry, generalized L-levels Holstein-Primakoff transformation, Hartree-Fock method and large-N saddle point approximation of the partition function, we extend the results obtained in [PRA 111, L020201 (2025)] to $L \geq 3$ and finite temperature $T>0$. 
In particular, we show that at $T=0$  for $U<U_c\sim -1/N$, the ground state is L-fold degenerate, 
while for positive temperatures, the critical temperature is both proportional to $N$ and $U$, i.e. $T_c \propto -U  N$, making this phase transition particularly suitable for large-N fermions.

\end{abstract}

\maketitle

\section{Intro}
The Fermi-Hubbard model stands as a cornerstone of modern condensed matter physics, providing a paradigmatic framework for understanding strong correlations, magnetism, and unconventional superconductivity \cite{Hubbard_1963,Gutzwiller_1963,Scalapino_2012,review_Arovas_2022,review_Corboz_2022}. 
Its SU($2$) incarnation is famously intractable in two dimensions and is central to theories describing the physics of cuprates\cite{Anderson_1987,Rice_1988}. 
Generalizing the spin symmetry from SU($2$) to SU($N$) offers a powerful theoretical extension, where the flavor number $N$ serves as a tunable control parameter\cite{Assaraf_1999,wu_exact_2003,Honerkamp2004,Capponi_annals_2016,Ibarra_Garcia_Padilla_2024,Chen_2024}. 
This generalization not only enriches the phase diagram with exotic magnetic and superconducting orders but also provides a unique pathway to analytical and numerical control, particularly in the large-$N$ limit\cite{affleck_exact_1986,Affleck_1988,Rokhsar_1990,Marder_1990,Read_1991,Chung_2001,Veillette_2007,Polychronakos_2024}.\\

The experimental realization of SU($N$) symmetric systems has seen remarkable progress in recent years, particularly through the use of ultracold alkaline-earth atoms in optical lattices\cite{Wu_review_2006,gorshkov_two_2010,Cazalilla_2014}. 
These atoms exhibit a natural decoupling of their nuclear spin from electronic degrees of freedom, enabling the creation of highly tunable SU($N$)-symmetric Fermi-Hubbard models (FHM) where $N$ can be as large as $10$\cite{takahashi2012,Pagano2014,Scazza2014,Zhang2014,Hofrichter_2016,Becker_2021,taie2020observation,Fallani_2022,Pasqualetti_2024}. 
The large-$N$ limit has become even more experimentally relevant with the recent proposal of using shielded ultracold molecules, where $N$ would reach up to $36$ \cite{Mukherjee_2024}.
The ultracold atoms allow precise control over hopping amplitudes $J$, interaction strengths $U$, and chemical potential.

Additionally, quantum simulation platforms such as dopant-based quantum dots in silicon \cite{Salfi_2016,Wang_2022} and optical tweezer arrays \cite{Wei_2024,Chew_2024} have successfully emulated extended Fermi-Hubbard physics. 
These systems provide complementary approaches for probing many-body correlations, thermodynamic properties, and quantum phase transitions in low-dimensional geometries, with recent experiments achieving unprecedented control over lattice parameters and single-site detection capabilities.
The flexibility of these experimental platforms in tuning symmetry, dimensionality, and interaction parameters makes them ideal for studying the finite-size effects and temperature-dependent phenomena central to this work.\\

An interesting aspect of SU($N$) models emerges in the context of small, finite systems. While traditional phase transitions require a thermodynamic limit in the number of lattice sites $L \rightarrow \infty$, an alternative limit can be constructed: for a fixed, small number of sites $L$, a large number of degenerate orbitals $N$ allows the system to contain many fermions while still accomodating the Pauli exclusion principle.
This can effectively play the role of a thermodynamic limit, stabilizing sharp phase transitions even in systems with very few sites. 
This phenomenon is illustrated in the simplest case of a two-site Hubbard chain (the "Hubbard dimer").
As established in \cite{Nataf_2025}, this model is Bethe-ansatz solvable for any $N$, and its spectrum admits an exact mapping to that of the Lipkin-Meshkov-Glick (LMG) model \cite{LMG_1965}. 
This mapping reveals a ground state quantum phase transition of second order at a critical interaction strength for attractive couplings.\\

\begin{figure*}
    \centering\includegraphics[width=1\linewidth]{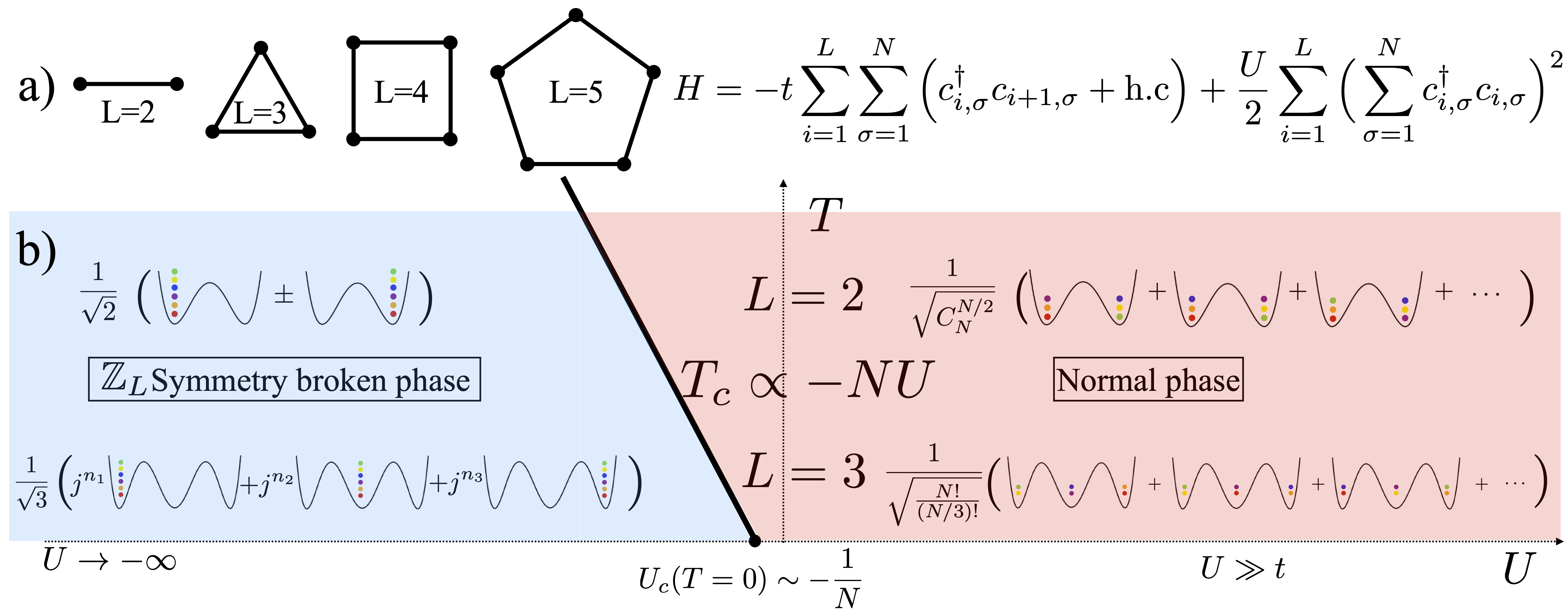}
    \caption{a) System and Hamiltonian under consideration: L sites ring and SU(N) Fermi-Hubbard model (FHM) with uniform hopping $t$ between nearest neighbors and on-site interaction $U$  b). Sketch of the phase diagram in the interaction/temperature $(U,T)$ plane and schematic depiction of the ground states for $L=2$ and $L=3$. In particular,  in the limit of large repulsive interaction $U \gg t$ (with fixed  $t>0$), the ground state is non-degenerate and its wave-function is uniformly distributed among the sites, while in the attractive limit $U/t \rightarrow -\infty$, the ground state manyfold is $L$ degenerate and the $\mathbb{Z}_L$ symmetry is broken in the thermodynamical limit. For a finite number $M$ of particles  (set to $N=M$ in our study), some linear superposition of quasi degenerate vacua {\it restore} the symmetry.  For instance, for  $L=3$, introducing $j \equiv \exp{2 i \pi /3}$, the sets of integers $\{n_1,n_2,n_3\}=\{0,0,0\},\,\{2,1,0\}, \,\{1,2,0\} $ define three orthonormal ground states which also diagonalize the $\mathbb{Z}_{L=3}$ parity operator (see Eq. \eqref{eq:parity}). At $T=0$, the critical interaction $U_c$ is such that $N U_c =-f(L)$ where $f$ is a decreasing function of $L$ for $L>2$ (cf Tab. \ref{tab:placeholder}).
    Finally, for $T>0$, we show in Sec. \ref{finite_temp} that the critical temperature $T_c$ is both proportional to $N$ and $-U$. }
    \label{fig:sketch}
\end{figure*}

In this work, we investigate the fate of this quantum critical point at finite temperature and for $L\geq 2$. 
While the $T=0$ transition in the $L=2$ dimer is well-understood, the finite-temperature phase diagram of few-sites SU($N$) Fermi-Hubbard chains remains unexplored. 
At $T = 0$, we systematically study chains of length $3 \leq L \leq 6$ for both large (finite and infinite) number of particles and colors  and for $T > 0$, we consider $L=2,3$ and $4$ sites, exploring how thermal fluctuations interplay with quantum correlations to produce, destroy, or modify the critical behavior. \\
%
%

The paper is organized as follows.
In Sec. \ref{sec:model}, we introduce the model in the parameters range we are interested in, the decomposition of the Hilbert space in independent sectors and some finite-N spectra. 
Then, in Sec. \ref{HP}, we show the Holstein-Primakoff (HP) representation of the SU(N) FHM which is particularly suited to the system under investigation. We calculate the eigen-modes in the small $U$ region (cf Sec. \ref{HP_weak}), and calculate the critical points in Sec. \ref{large_N_minim} through the minimization of a large-N energy functional.
In Sec. \ref{QPT}, the order parameter of the transition associated with the $\mathbb{Z}_L$ symmetry breaking is derived and reveals different orders of the transition depending on $L$.
Then,  in Sec. \ref{finite_temp}, we study the finite temperature partition function and from a large-N saddle point treatment, we obtain the (linear) behavior of the critical temperature $T_c$ as a function of $N$ and $U$.
Finally, conclusions and perspectives are drawn.
\section{Model and Spectra}
\label{sec:model}

The SU($N$)-invariant FHM Hamiltonian $H$ can be written as the sum of a kinetic Hamiltonian $H_K$ and an interaction Hamiltonian $H_I$, i.e. $H=-t H_K+\frac{U}{2} H_I,$ with:
\begin{align}
\label{eq:Ham} 
    H_K =  \sum\limits_{i} \left(E_{i, i+1} + h.c. \right)  \hspace{.5cm} \text{and} \hspace{.5cm} H_I = \sum\limits_{i=1}^L E_{i, i}^2,
   \end{align}
where the SU(N) invariant hopping terms read:
\begin{equation}
    E_{i, j} =  \sum\limits_{\sigma=1}^{N} c^\dagger_{i, \sigma} c_{j, \sigma}.
\end{equation}
$c$ (resp. $c^\dagger$) are fermionic annihilation (resp. creation) operators, $1 \leq \sigma \leq N$ labels the pseudospin degree of freedom, and the latin indices $i=1 \cdots L$ stand for the site indices.
As shown in Fig. \ref{fig:sketch}, we consider small chains of length $L$ with periodic boundary conditions, with uniform hopping amplitude $t$ and on-site interaction $U$, but our approach can be readily extended to more general set of parameters.\\
The operators $E_{i, j}$ and $E_{k,l}$ (for $i,j,k,l=1 \cdots L$),  satisfy the commutation relations of the generators of the Lie algebra of the unitary group $U(L)$:
\begin{equation}
\label{commutation}
[E_{ij},E_{kl}]=\delta_{jk}E_{il}-\delta_{li}E_{kj}.
\end{equation}

The SU(N) FHM can therefore be analyzed using the Lie algebra representation theory of the unitary group $U(L)$ \cite{Gelfand_1950}.
This provides an efficient framework for implementing the full SU(N) symmetry in the exact diagonalization (ED) of the model~\cite{Botzung_2023_PRL,Botzung_2023}.
This approach, which should be seen as the analogue of using the representation theory of the permutation algebra for the Heisenberg SU(N) Hamiltonian \cite{nataf2014},
is briefly summarized below.\\

\begin{figure}[h]
    \centering\includegraphics[width=1\linewidth]{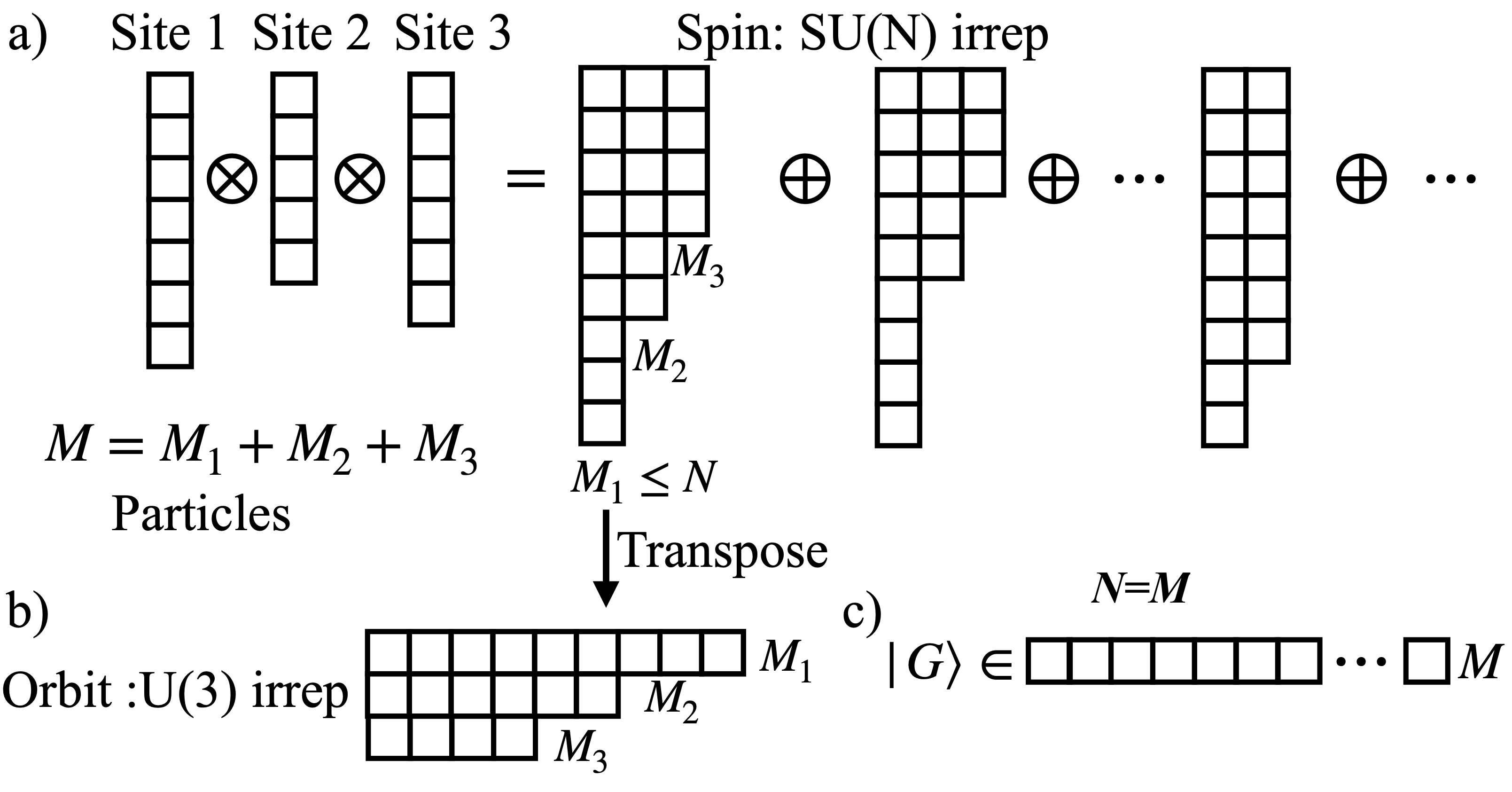}
    \caption{a)Relevant Young Diagrams (YDs) $\alpha$ for the SU(N) Fermi-Hubbard model on $L$ sites ($L=3$ here): at most $L$ columns and $N$ rows.
    For each YD, the total number of boxes is equal to the number of particles that we note $M$. b) The orbital degrees of freedom live in the $U(3)$ irrep represented by the {\it transposed} YD $\bar{\alpha}$. c) When $M=N$, the ground state $\vert G \rangle$ lives in the fully symmetric U(L) irrep $\bar{\alpha}=[N]$.}
    \label{fig:irrep}
\end{figure}

The irreducible representations (irreps) of SU(N) are labelled by Young Diagrams (YDs) of shape $\alpha = (\alpha_1, \alpha_2..., \alpha_N)$, where $\alpha_j$ is the number of boxes on the $j^{\text{th}}$ row (for $1 \leq j \leq N$), satisfying $\alpha_j \geq \alpha_{j+1} \geq 0$ (cf Fig. \ref{fig:irrep}). 
We also define $\bar{\alpha}$, the {\it transposed} shape of $\alpha$, i.e. the shape whose rows have length $M_1, M_2, \cdots M_L$ which are the length of the columns of $\alpha$
(cf Fig. \ref{fig:irrep} b)).
We call $M$ the number of fermions in the system and it is equal to the total number of boxes, i.e. $M=\sum_{i=1}^N \alpha_i=\sum_{i=1}^L M_i$.
For a fermionic wavefunction whose color (magnetic) degrees of freedom have the symmetry of the SU(N) irrep $\alpha$, the global antisymmetry condition imposes that its orbital degrees of freedom (which are impacted by the hopping operators $E_{i,j}$ for $1 \leq i,j \leq L$) live in the irrep $\overline{\alpha}$ of U(L).
For instance, if the magnetic degrees of freedom of the wave-function are fully antisymmetric (i.e. living in a one column SU(N) irrep), then its orbital degrees of freedom should be fully symmetric (i.e. living in the one-row YD). More generally, the YDs of N-colors fermions on L-sites, should not contain more than $L$ columns and $N$ rows,
as it appears when we tensor product L-times (following the Littlewood Richardson rules \cite{itzykson}) the one-column irreps corresponding to the fully antisymmetric fermionic wave-function of each site.

Unless otherwise specified, we restrict our study throughout this paper to a fixed number of particles $M=N \gg L$.

For fixed $M$, the Hilbert space decomposes into a set of invariant sectors, each of them being the direct sum of $D^N_\alpha$ independent and equivalent  $U(L)$ irrep $\overline{\alpha}$, where $D^N_\alpha$ is the dimension of the SU(N) irrep $\alpha$ \cite{Botzung_2023_PRL}.

An appropriate basis for the corresponding $U(L)$ irrep $\overline{\alpha}$ is the set of the $D^L_{\overline{\alpha}}$ semi-standard Young tableaux (SSYT) of shape $\overline{\alpha}$, i.e.  filled with numbers from $1$ to $L$, ascending from left to right (repetitions allowed), and strictly ascending from top to bottom.
Then, the dimension of the Hilbert space for fixed $M$ is $\sum_{\alpha} D^N_\alpha D^L_{\overline{\alpha}}$, where the sum runs over all the YDs $\alpha$ of $M$ boxes, with less than $N$
rows and $L$ columns \cite{Botzung_2023_PRL}.
The matrix elements corresponding to the operators $E_{i, j}$ can be easily computed following the formulas in Ref.~\cite{Botzung_2023_PRL}.
We use standard ED techniques, such as e.g. Lanczos-based methods for the spectrum and observables at $T=0$ (cf Sec. \ref{sec:model}, Sec. \ref{HP} and Sec. \ref{QPT}) and the full diagonalization for the exact partition function at $T>0$  (cf Sec. \ref{finite_temp}) to compute the eigenvalues and eigenvectors of the Hamiltonian, and each eigen-level is {\it a priori} $D^N_\alpha$-degenerate (not accounting for supplementary spatial symmetry).
%

\begin{figure}
    \centering\includegraphics[width=1\linewidth]{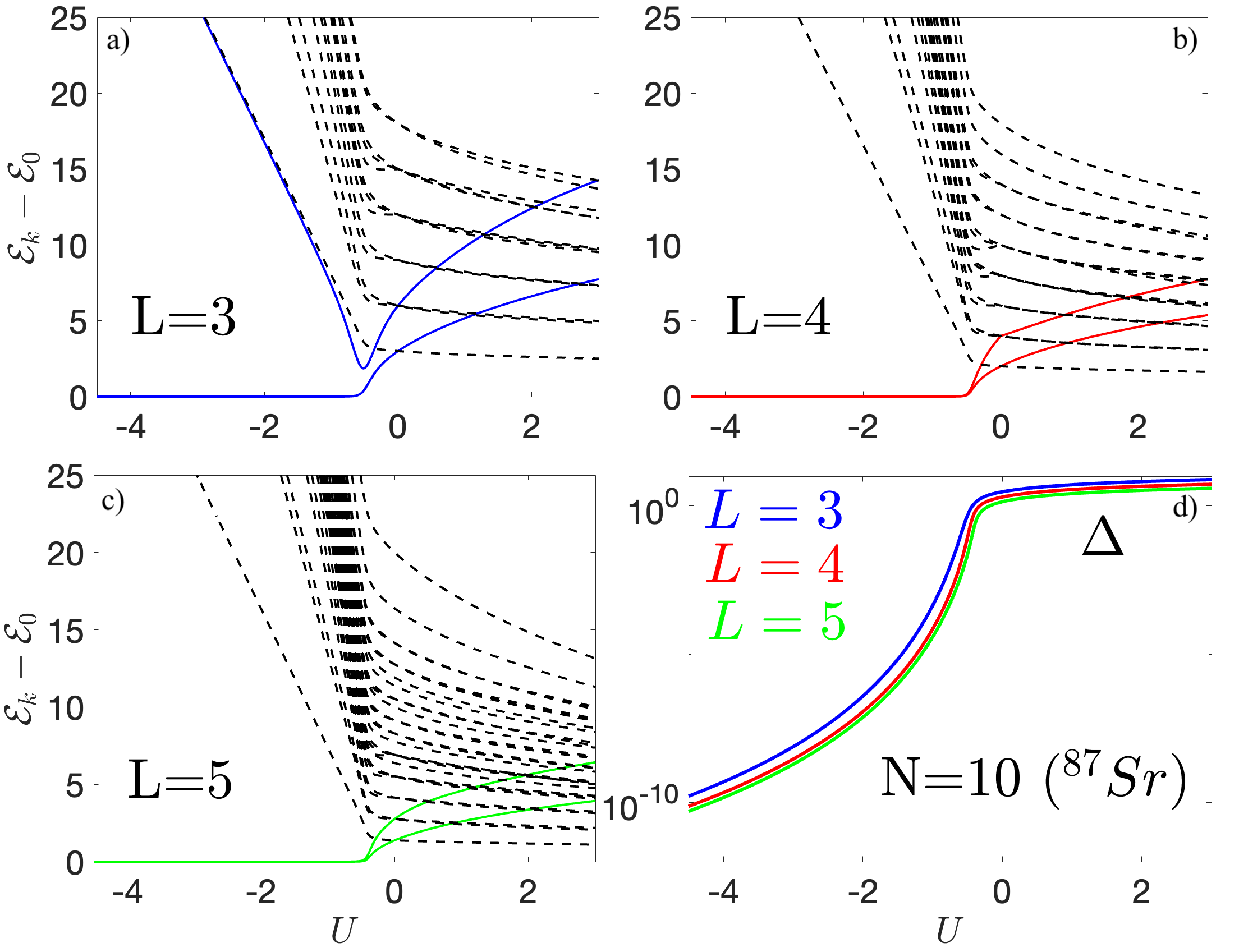}
    \caption{ For $N=10$ flavors (relevant for the atom $^{87}$Sr \cite{Stellmer_2013,Zhang2014,Bataille_2020,Ahmed_2025}), we plot the spectra for the periodic chain with $L$ sites, for L=3 (a), L=4 (b) and L=5 (c) as a function of the on site interaction $U$ for $t=1$. The energies plotted in solid lines are the eigen-energies within the SU(10) singlet sector (corresponding to the fully symmetric $M=N=10$ boxes  irrep for the obital degrees of freedom), while the dashed lines are minimal energies from other irreps. Ground state energy $\mathcal{E}_0$, which always lives in the SU(10) irrep, is withdrawn. d)  Singlet gaps $\mathcal{E}_1-\mathcal{E}_0$ for $L=3,4,5$: for attractive $U$, it decreases exponentially with $\vert U \vert$, signaling the finite-size version of the quantum phase transition, like for $L=2$ \cite{Newman_1977,Nataf_2025}.}
    \label{fig:energy_N10}
\end{figure}

This formalism allows for a drastic reduction of the effective dimension of the matrices to diagonalize,
especially in the considered situation, i.e.  when $N \gg L$.
As a typical example, for $L = 3$ sites and $N = 42$ colors, the full Hilbert space is of dimension $2^{NL}=2^{126}$.
The sector with $M = N$ atoms is of dimension $\approx 5 \times 10^{33}$.
Taking into account the color conservation reduces the largest effective Hilbert space down to $\approx  10^{20}$.
On the other hand, the largest matrix we need to consider to study this sector is only of dimension $4375$ (out of the $169$ possible irreps).

For $N=10$ colors, relevant for the $^{87}$ Sr cold atoms \cite{Stellmer_2013,Zhang2014,Bataille_2020,Ahmed_2025}, we show the eigen-energies $\mathcal{E}_k$ for $L=3,4$ and $5$ in Fig. \ref{fig:energy_N10} for $t=1$ as a function of $U$: in particular, the ground state (of energy $\mathcal{E}_0$) is always a SU(10) singlet, i.e. its (orbital) wave-function lives in the fully symmetric U(L) one row N-boxes YD (cf also Fig. \ref{fig:irrep} c).
This property, which is general for the parameters and the filling under investigation (i.e for $M=N$), will have important and fortunate consequences about the accuracy and the relevancy of the HP transformation in the Sec. \ref{HP}.

Secondly, the {\it singlets} gap (defined as the difference between the first singlet excited eigen-energy and the ground state energy) shown in Fig. \ref{fig:energy_N10} d), exhibits
an exponentially decreasing behavior while going into the attractive region of the interaction $U$.
This feature is reminiscent of the $L=2$ case \cite{Newman_1977,Nataf_2025} and is compatible with the occurence of a QPT in the thermodynamical limit ($N=M \rightarrow \infty $, L fixed) separating a symmetry broken phase (in the region $U<0$) from a normal phase (in the region $U>0$). We address such a limit thanks to the Holstein-Primakoff transformation in the next section.

\begin{figure}
    \centering\includegraphics[width=1\linewidth]{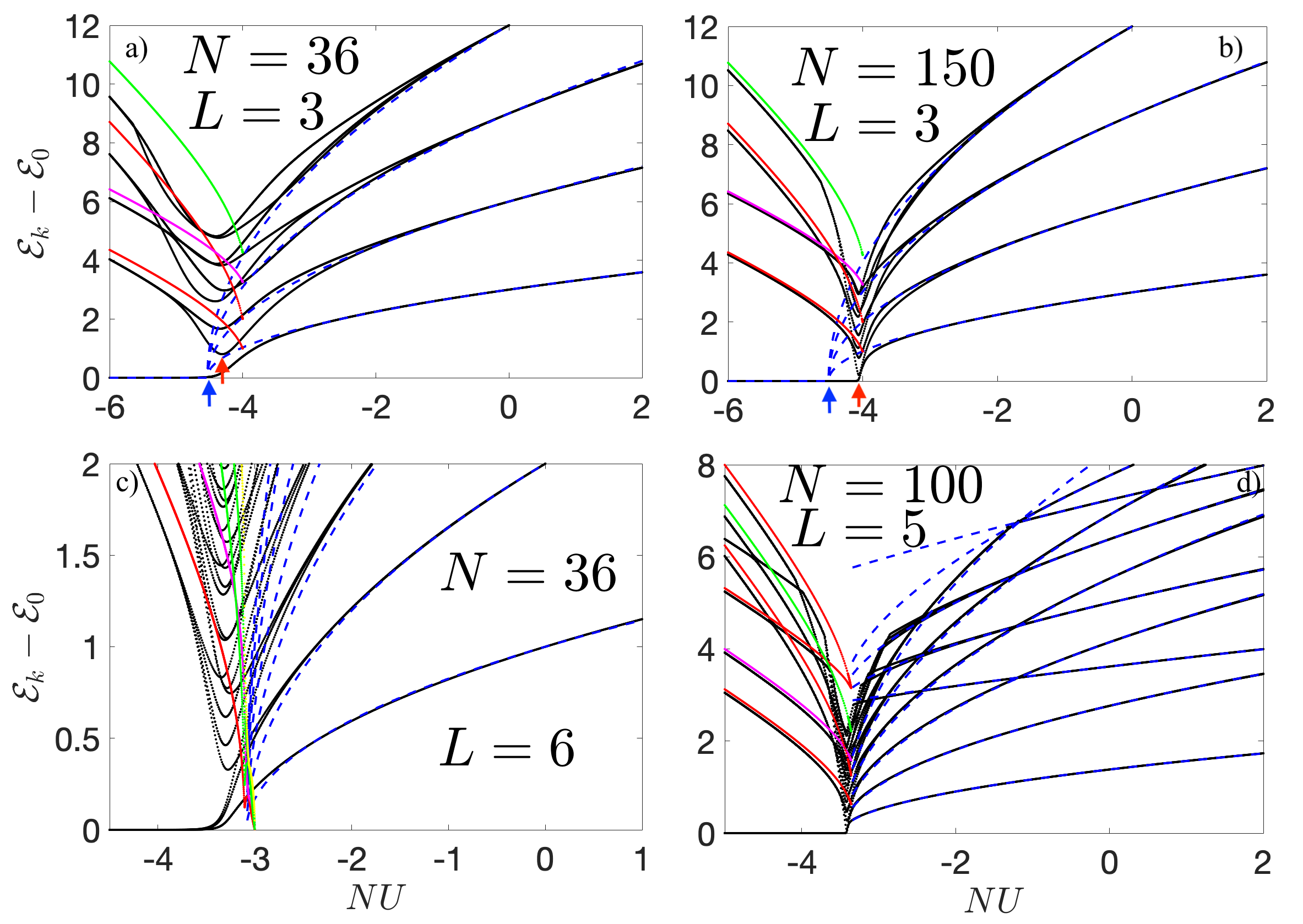}
    \caption{ Spectra of the SU(N) FHM on L sites for $t=1$ for various values of N and L. 
     In black, ED results within the singlets irreps for $M=N$ particles. We show in colored lines the large-N pulsation frequencies obtained from the Holstein-Primakoff transformation: in dashed blue, for the normal phase (corresponding to the weak U region):  For  a) and b),  $q \omega_{k=1}^N$ for $q=1,2,3$ and $4$ and $L=3$ ; For c) and d), we add other integer linear combination of the $\omega_{k}^N$  (cf Eq. \eqref{HP_pulsation}). There is an offset between the value $NU^{k=1}_c$ where $\omega_{k=1}^N$ vanishes (cf Eq. \eqref{wrong_UC}), shown with the vertical blue arrow in a) and b), and the (finite-N) value $NU_c^N$, where  the $L^{\text{th}}$ gap admits a minimum (see text for details), shown in vertical red arrow. 
The symmetry broken phase frequencies are displayed in non-blue solid lines.
   \label{fig:pulsation}}
\end{figure}

\section{Large-N solution through Holstein-Primakoff Transformation}\label{HP}
\subsection{Weak U region} \label{HP_weak}
The L-levels Holstein-Primakoff (HP) transformation is an exact bosonic representation of the fully symmetric $U(L)$ irrep \cite{HP_1940,Papanicolaou_1984}.
As such, we expect to obtain an accurate approximation of the ground state and of the (singlet) gaps in the large-$N$-development of the HP representation.

To provide a harmonic approximation of the spectrum in the weak $U$ region, one should start from the diagonalization of
 the (ring-like) kinetic part of the Hamiltonian (cf Eq. \eqref{eq:Ham}):
\begin{equation}
-tH_K=-t\sum_{i} E_{i,i+1}+h.c=-2t \sum_{k=1}^{L} \cos{\Big{(}\frac{2 k \pi}{L}\Big{)}} \tilde{E}_{k,k}, \label{Eq:H_K}
\end{equation}
where the rotated Lie generators $\tilde{E}_{i,j}$  ($i,j=1 \cdots L$), which also satisfy the commutation relations Eq. \eqref{commutation}, are function of the unrotated ones $E_{l,s}$ ($l,s=1 \cdots L$)
through:
\begin{equation}
\tilde{E}_{p,j} = \sum_{l,s}C(p,l)^*C(j,s) E_{l,s}.
\end{equation}
The $k^{th}$ column of the $L \times L$ Fourier matrix $C$ (for $k=1 \cdots L$) is the $k^{th}$ eigenvector of the $L \times L$  adjacency matrix representing the lattice,
i.e. $C(j,k)=1/\sqrt{L} e^{ikj}$.
This is nothing but band theory expressed in terms of the $U(L)$ generators.
In the rotated basis, the interaction part  of the Hamiltonian, i.e. $H_I=\sum\limits_{i} E_{i, i}^2$ is equal to:
\begin{equation}
 H_I=\sum_{k,l} \sum_{r,s} \Big{(} \sum_i C(i,k)^*C(i,l) C(i,r)^*C(i,s) \Big{)} \tilde{E}_{k,l} \tilde{E}_{r,s}. \label{Eq:H_I}
\end{equation}
The band $k=L$ of minimal energy $-2t$ (for $t>0$) should be taken as the referring level in the HP representation of the generators of $U(L)$ in the fully symmetric $N=M$ boxes irrep \cite{Papanicolaou_1984,Katriel_1983,papanicolaou1988,Molmer_2010,Romhanyi_2012}:
\begin{align}
\tilde{E}_{i,j} &= a^{\dag}_i a_ j \hspace{2.5cm}  \text{for} \hspace{.5cm} 1 \leq i,j \leq L-1 \nonumber \\
\tilde{E}_{i,L} &= a^{\dag}_i \sqrt{N-\sum_{j<L} a_j^{\dag} a_j}\hspace{.5cm}  \text{for} \hspace{.5cm} 1 \leq i \leq L-1 \nonumber \\
\tilde{E}_{L,L} &=N-\sum_{j<L} a_j^{\dag} a_j, \label{HP_transformation}
\end{align}
where we have introduced $L-1$ pairs of creation and annihilation bosonic operators $a_j^{\dagger}$ and $a_j$ ($j=1 \cdots L-1$).
The  Hamiltonian becomes:
\begin{multline}
    H \simeq 2t \sum_{k=1}^{L-1} (1- \cos{\Big{(}\frac{2 k \pi}{L}\Big{)}})a_k^{\dagger} a_k \\
    + \frac{UN}{2L} (2\sum\limits_{k} a^\dagger_k a_k + \sum\limits_{k} (a^\dagger_k a^\dagger_{-k}  + a_k a_{-k}) ),
\end{multline}

where we dropped the constants and where we considered the harmonic and large-N limit for $H_I$. In particular, we selected in Eq. \eqref{Eq:H_I} only the terms involving twice the index $L$.
 Finally, a Bogoliubov transformation implying the diagonalization of $L-1$ independent $2 \times 2$ matrices leads to the normal frequencies in the normal phase:
 \begin{align}
 \omega_k^N = \sqrt{4t \sin^2 \frac{k\pi}{L}   \times \left( 4t \sin^2 \frac{k\pi}{L} + \frac{2UN}{L} \right)}. \label{HP_pulsation}
 \end{align}

For $k=1 \cdots $ Floor($(L-1)/2$), the pulsation is twice degenerate and for $L$ even, there is an additional non-degenerate frequency corresponding to $k \equiv L/2$.

In Fig. \ref{fig:pulsation}, the singlet gaps for various values of $N$ and $L\geq 3$ are displayed as a function of $NU$ and exhibit very good agreements for small interaction $\vert U \vert$ with the integer combination of the HP pulsations (cf Eq. \eqref{HP_pulsation}), i.e. quantities written as $\sum_k \sum_{n_{q_k} \in \mathbb{N}} n_{q_k} \omega_k^N$, which describe the full energy spectrum in a multi-boson quadratic model.
At fixed $t>0$, the critical value $U^{k=1}_c$ is equal to the smallest (in absolute value) $U$ for which one of the $\omega_k$ vanishes. It corresponds to $k=1$ so that
\begin{equation}
U^{k=1}_c/t = -\frac{2 L}{N}\sin^2( \pi /L). \label{wrong_UC}
\end{equation}
Note that this estimation has been derived at first order in $1/N$ and for $L \geq 3$. In particular, for $L=2$, the above result for $U^{k=1}_c$ would be twice as big as in Ref. \onlinecite{Nataf_2025},
 due to the double counting of the hopping for $L=2$ with periodic boundary conditions.
 
 To test Eq. \eqref{wrong_UC} for $L \geq 3$, one can introduce $U_c^N$, some finite-{\it size} (i.e. particles or N since $N=M$) version of the critical value $U$, defined as the value of $U$ at which the $L^{\text{th}}$ singlet gap (defined as $\mathcal{E}_{L}-\mathcal{E}_0$) admits a minimum (cf Fig. \ref{HP_pulsation} a) and b) where $NU_c^N$ appears as a small vertical red arrow for $L=3$).
While the mismatch between $NU^{k=1}_c/t=-4.5$ (cf Eq. \eqref{wrong_UC}  for $L=3$, and appearing as a small vertical blue arrow in Fig. \ref{fig:pulsation} a) and b)) and  $NU_c^N/t \simeq -4.31 $  for $N=36$ (relevant for cold molecules Na$^{40}$K  \cite{Mukherjee_2024}) could be attributed to finite-N  effect, it is clear on Fig. \ref{fig:pulsation} b), where $N=150$,  that this is not true, since there $NU_c^N/t \simeq -4.06 $, which is even further from $-4.5$.
To understand such a discrepancy, one should introduce the possibility of macroscopically {\it displaced} bosonic modes in the HP transformation, i.e.
$\langle a_j \rangle, \langle a_j^{\dagger} \rangle \sim \sqrt{N}$, for $j=1 \cdots L-1$ in Eq. \eqref{HP_transformation}.

\subsection{Classical Minimization in the large-N limit}
\label{large_N_minim}
We can {\it priori} assume that
\begin{align}
 a_j = \sqrt{N} \mu_j + \delta_j, \label{eq:displacement}
\end{align}
 where the $\mu_j$ are  c-numbers of order $1$, and where the creation and annihilation operators $\delta^\dagger_j, \delta_j$ are also bosonic for $j=1 \cdots L-1$.

In frustrated magnetism, such a macroscopic displacement corresponds to a global rotation towards a different classical ground state in the large-N limit, around which quantum fluctuations
could also be taken into account through spin-wave or linear flavor wave theories \cite{papanicolaou1988,Marder_1990,joshi1999,toth2010,Romhanyi_2012,Kim_2017}.
In quantum optics, the quantities $\mu_j$ are usually named {\it coherences} \cite{Hayn_2011,Nataf_2012,Baksic_2013,Jiang_2025}, and the $U(L)$ {\it coherent} states within the fully symmetric irrep can be built from the HP representation \cite{Gilmore_1979,Katriel_1983}.

After having introduced $\mu_L\geq 0$ defined by $\mu_L^2 = 1 - \sum\limits_{k\neq 0} \vert \mu_k\vert^2$ and implemented Eq. \eqref{eq:displacement} into Eqs. \eqref{HP_transformation} for the unrotated $E_{i,j}$ \footnote{Since adding coherences is equivalent to a global rotation in the large-N limit, one can directly start from the unrotated Hamiltonian operators.}, the energy (per particle or color) functional $E(\vec{\mu})$ associated with the SU(N) FHM on the L-sites ring (cf Eq. \eqref{eq:Ham}) reads (to lowest order in $N$):
\begin{equation}
    E(\{\mu_k \}) = \frac{UN}{2} \sum\limits_{k=1}^{L} \vert \mu_k\vert^4  -t \sum\limits_{k=1}^{L} (\mu_k^* \mu_{k+1} + \text{h.c}), \label{eq:functional}
\end{equation}
(where $\mu_{L+1}\equiv \mu_1$).

Looking for the bests $\{ \mu_j, j=1 \cdots L-1\}$ gives then raise to a simple classical minimization problem depending on the parameters $L$ and $NU$.

\begin{figure}
    \centering\includegraphics[width=1\linewidth]{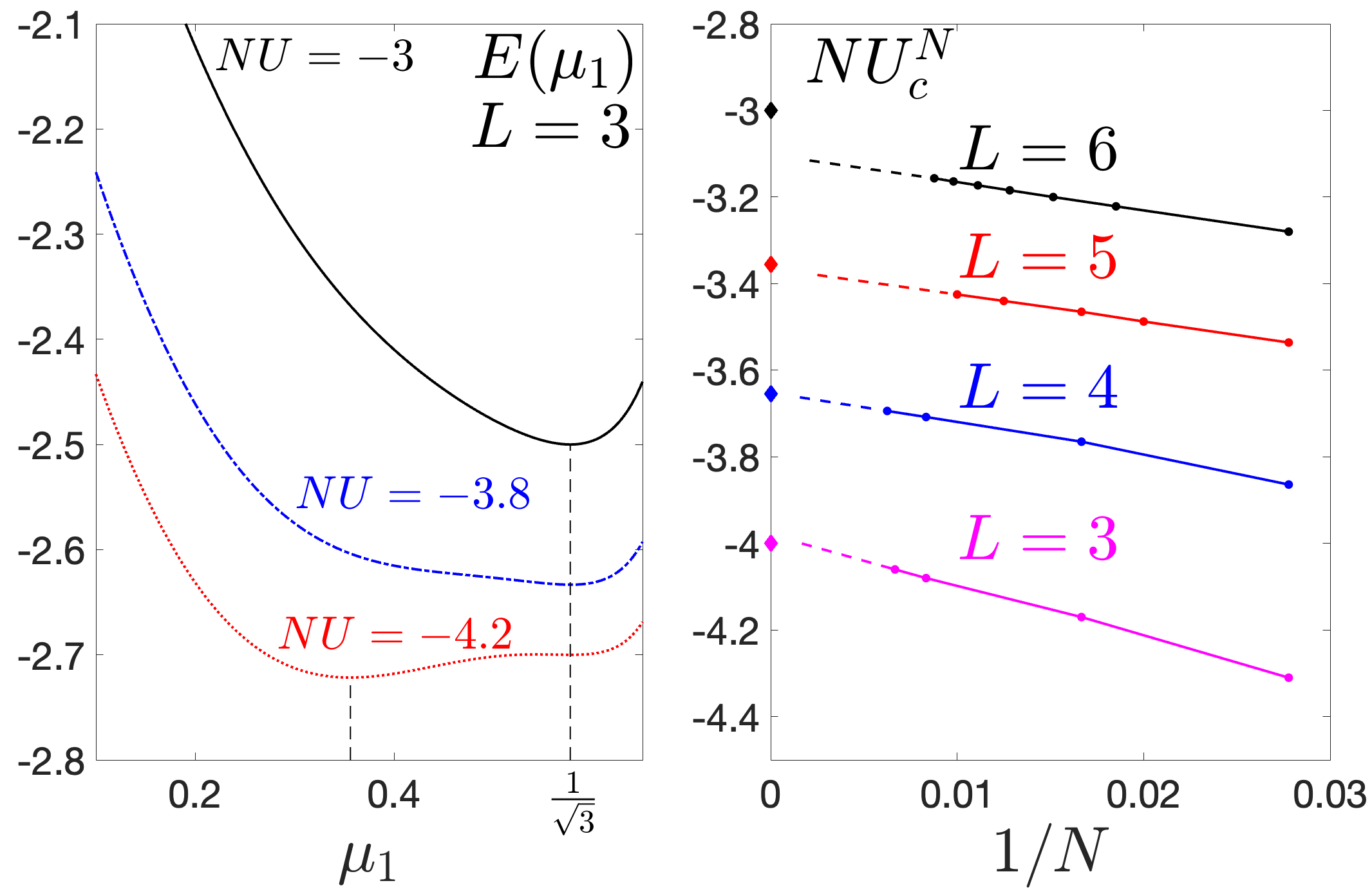}
    \caption{a) Energy per site $E(\mu_1) $ functional in the infinite $N$ limit for $L=3$, at $t=1$, as a function of the macroscopic field displacement $\mu_1$ (cf Eq. \eqref{eq:functional_L3_mu1}) for various values of $NU$: while $E(\mu_1)$ has a global minimum at $\mu_1=1/\sqrt{3}$ for small $\vert U\vert$, when $NU<N U^{\infty}_c=-4.00$, a new global minimum appears suddenly for finite value of $\mu_1 \equiv \mu_1^m$. b) For $L=3,4,5$ and $6$, comparison between $N U^{\infty}_c$ (shown as colored diamonds, extracted from Tab. \ref{tab:placeholder}) and $NU_c^N$ (shown as points and colored lines, and defined as the locations of the minimum of the L$^\text{th}$ singlet gap, i.e. $\mathcal{E}_L-\mathcal{E}_0$ cf also Fig. \ref{fig:pulsation} a) and b)), as a function of $1/N$. Dashed lines are linear fitting from the last points. \label{fig:minim} }
\end{figure}


Let's first focus on $L = 3$: from inversion symmetry, one can impose $\vert \mu_1\vert = \vert \mu_2\vert$. One can also numerically check that global minimal solutions
correspond to real positive $\mu_j$ (for $j=1,2$ and $3$) so that we are left with the (easy to plot and analyze) one-parameter functional:
\begin{equation}
    E(\mu_1) = -2t\left(\mu_1^2 +2 \mu_1 \sqrt{1-2 \mu_1^2}\right)+ \frac{UN}{2} \left(1-4 \mu_1^2+6 \mu_1^4 \right). \label{eq:functional_L3_mu1}
\end{equation}

While $\mu_1=1/\sqrt{3}$ is always a local minimum (it nullifies $dE(\mu_1)/d \mu_1$), we show on Fig. \ref{fig:minim} a) that when $NU/t < NU_c^{\infty}/t=-4.00$, a new global minimum appears for finite $\mu_1 \equiv \mu_1^m$, where $0<\mu_1^m<1/\sqrt{3}$. 
In particular, when $NU/t \rightarrow -4^{-}$, the location of this new global minimum $\mu_1^m$ is not arbitrarily close to $1/\sqrt{3}$.
The finite offset impacts the nature of the transition, as we will see in the next section (i.e. being first order contrarily to the $L=2$ transition \cite{Nataf_2025}).

\begin{table}[]
    \centering
    \begin{tabular}{c|c|c|c|c|c|c|c|c|c|c|c}
        $L$ & 2 & 3 & 4 & 5 & 6& 8 & 10 & 12\\
        \hline
        $NU^{\infty}_c$ & $-2$ & $-4.00$ & $-3.65$ & $-3.36$ & $-3.00$&$-2.34$&$-1.91$& $-1.61$
    \end{tabular}
    \caption{Critical value of the interaction, i.e. $NU^{\infty}_c \equiv NU_c$ ($t=1$),  in the thermodynamical limit $N=M=\infty$ at $T = 0$, according to the coherent state approach. For $L=2$, the value is analytical \cite{Nataf_2025}, while for $L>2$, it is obtained from the numerical minimization of Eq. \eqref{eq:functional} (up to $10^{-2}$).}
    \label{tab:placeholder}
\end{table}

In Tab. \ref{tab:placeholder}, we give the values of $NU_c^{\infty}$ that we obtain for $L=3,4, \cdots 12$ from a numerical minimization of Eq. \eqref{eq:functional} ,
and we have added the $L=2$ analytical value ($NU_c=-2$, cf Eq. (19) in \cite{Nataf_2025}).
Interestingly, we realized that $U_c^{\infty} \simeq U_c^{k=1}$ (cf Eq. \eqref{wrong_UC}) only for $L\geq 6$.
In Fig. \ref{fig:minim} b), we compare $U_c^{\infty}$ to $U_c^N$ for $L=3,4,5$ and $6$ for different $N$ (up to  $N=120$ for $N=6$).
The agreement is very good: for $L=3,4$ and $5$, the linear fitting of $U_c^N$ as a function of $1/N$ (materialized by straight dashed lines in Fig. \ref{fig:minim} b)), conveys to an estimate
of $NU_c^{\infty}$ within less than 0.25 \% of the tabulated values in Tab. \ref{tab:placeholder}. For $L=6$, the matching is less good, only $3.3\%$, as the ED results point towards $-3.10$ and not $-3.00$. An inflection might occur for values of N larger than those considered here, but since the dimension of the matrix to diagonalize (i.e. $D_{\overline{\alpha}}^{6}$) is already $\approx 255 \times 10^6$ for $N=120$, this question is difficult to address through ED.


Finally, incorporating the quantum fluctuations through quadratic bosonic terms in the broken symmetry phase, one can also obtain the HP frequencies
for $U<U_c$, as computed in Appendix \ref{app:HP}, and shown in Fig. \ref{fig:pulsation}. Like for $U>U_c$, agreement with finite-N ED results improve with N, as seen when one compares Fig. \ref{fig:pulsation} b) (i.e for $L=3$ and $N=150$) to  Fig. \ref{fig:pulsation} a) (i.e. for $L=3$ and $N=36$).

\section{$T = 0$ QPT}
\label{QPT}
%
Thus, the SU(N) FHM on the L-sites ring for $M=N$ particles undergoes a Quantum Phase Transition (QPT) for fixed t at $U=U_c^{\infty} \equiv U_c$.
For $U > U_c$, the kinetic energy dominates, the atomic density is uniform over the sites and for 
$U < U_c$, the ground state manyfold is L-times degenerate and the attractive potential tends to condense all fermions on a single site. For finite N,
the first L eigenvectors are quasi degenerated with exponentially small gaps, and Schr{\"o}dinger-cat like forms of the wave-functions, as sketched on Fig. \ref{fig:sketch} for $L=2$ and $3$.
In particular, for $U < U_c$, the symmetry $\mathbb{Z}_L$ is broken. The explicit expression of the associated operator  in the $U(L)$ algebra reads:
\begin{equation}
\mathbb{Z}_L \equiv e^{ \sum_{j,k} \mathcal{M}_L(j,k)E_{j,k}}, \hspace{.25cm} \text{with}  \hspace{.25cm} \mathcal{M}_L \equiv \text{Log} \left ( T_L\right),\label{eq:parity}
\end{equation}
where the $L \times L$ matrix $T_L$ is the translation operator for the lattice. For instance, for $L=3$:
\begin{equation}
T_L=\begin{pmatrix}
0 & 1 & 0 \\0 &0 & 1 \\ 1 & 0 & 0 
\end{pmatrix} \Rightarrow 
\mathcal{M}_L= -\frac{2 \pi}{3 \sqrt{3}}\begin{pmatrix}
0 & -1 & 1 \\1 &0 & -1 \\ -1 & 1 & 0 
\end{pmatrix}.
\end{equation}

While these features for the broken-symmetry phase for $L>2$ mirror those of the $L=2$ model \cite{Nataf_2025}, the nature of the transition is different, as we will show by focusing on the most natural order parameter of the transition:

\begin{equation}
    \mathcal{O}_L =\frac{1}{LN^2} \sum\limits_{j = 1}^L \langle E_{j,j}^2\rangle - \frac{1}{L^2}. \label{eq:order_parameter}
\end{equation}

\begin{figure}[ht!]
    \centering\includegraphics[width=0.9\linewidth]{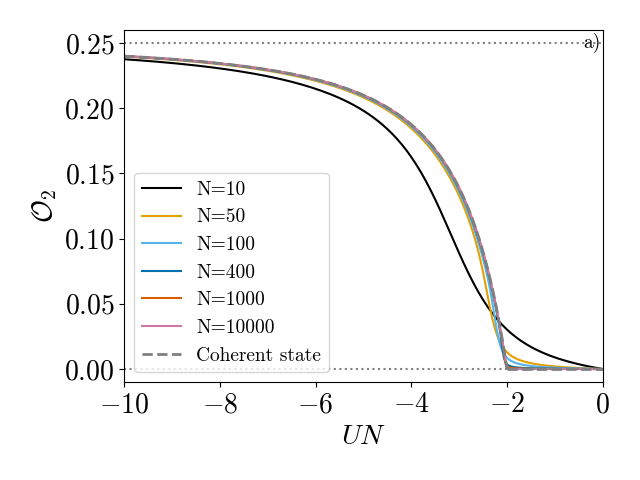}\\
    \centering\includegraphics[width=0.9\linewidth]{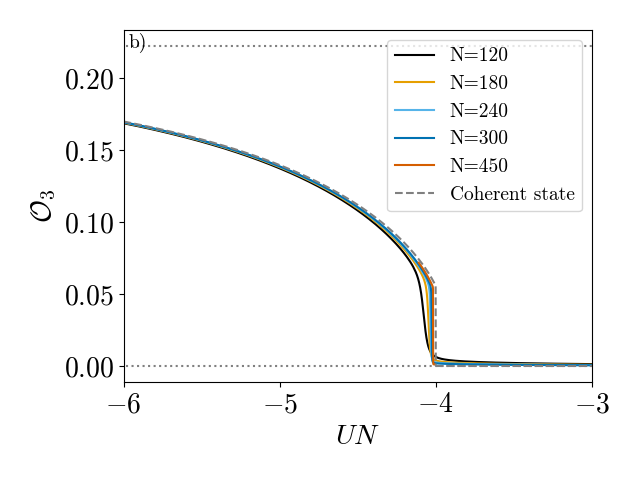}\\
    \centering\includegraphics[width=0.9\linewidth]{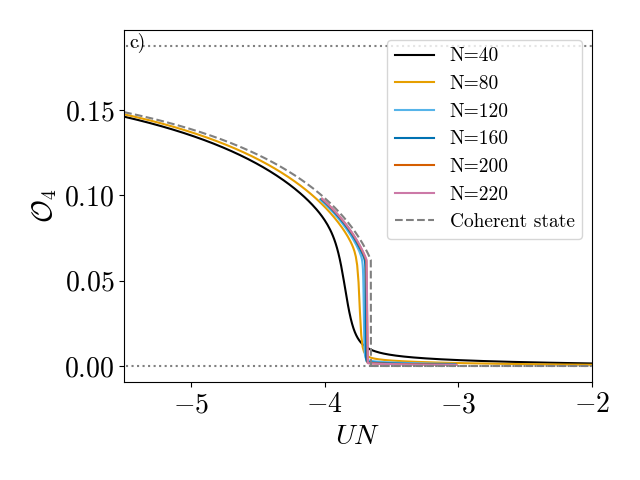}
    \caption{Order parameter $\mathcal{O}_L$ (cf Eq. \eqref{eq:order_parameter}) in the groundstate of the a) $L=2$, b) $L=3$ and c) $L=4$ SU(N) FHM on the L-sites ring (cf Eq. \eqref{eq:Ham}) at filling $M = N$ for $t=1$. In dashed-grey line, the thermodynamical limit is achieved thanks to a coherent-state approach (or large-N Holstein-Primakoff development with macroscopic coherences cf Eq. \eqref{eq:displacement} and \eqref{HP_transformation}). In color lines, ED results for various N. For $L = 2$, it predicts a second-order transition at $U_c N = -2$, while for $L=3$ and $L=4$ the transition becomes first-order, as shown in Fig. \ref{fig:OP0T-zoom} which is the zoomed-in view of the current figure. The dotted horizontal lines denote the $U = +\infty$ and $U = -\infty$ values of the order parameter. Cf text for details.}
    \label{fig:OP0T}
\end{figure}

 $\mathcal{O}_L$ is the average number of on-site pair per site \cite{Ibarra_2021}.
 It is experimentally relevant as the photoassociation (PA) process \cite{takahashi2012} which transfers the atoms in doubly occupied lattice
sites into highly excited molecular states is a direct measurement of doublons, and hence of the Mott insulating nature of the SU(N) fermionic systems on lattices\cite{taie2020observation,Fallani_2022,Pasqualetti_2024}.

In the large-N limit, the coherent states based approach developed in the previous section, directly gives:
\begin{equation}
\mathcal{O}_L =\frac{1}{L}  \left ( \sum_{i=1}^{L} \vert \mu_i^m \vert ^4 - \frac{1}{L}\right),
\end{equation}
where $\{ \mu_j^m, j=1 \cdots L \}$ minimizes $E(\vec\mu)$ (cf previous section).
In particular, for $L=3$, $\mathcal{O}_{L=3}=\frac{1}{3}(2/3+6(\mu^m_1)^4-4(\mu^m_1)^2)$, and more generally, for any $L$,
$\mathcal{O}_L  \rightarrow 0$ for $U/t \gg 1$ and $\mathcal{O}_L  \rightarrow 1/L-1/L^2$ for  $U/t \ll -1$.
 In Fig. \ref{fig:OP0T} we display $\mathcal{O}_L$ in the  thermodynamical limit and for finite N  for $L=2,3$ and $4$, 
showing a good convergence.

\begin{figure}[ht!]
    \centering\includegraphics[width=0.9\linewidth]{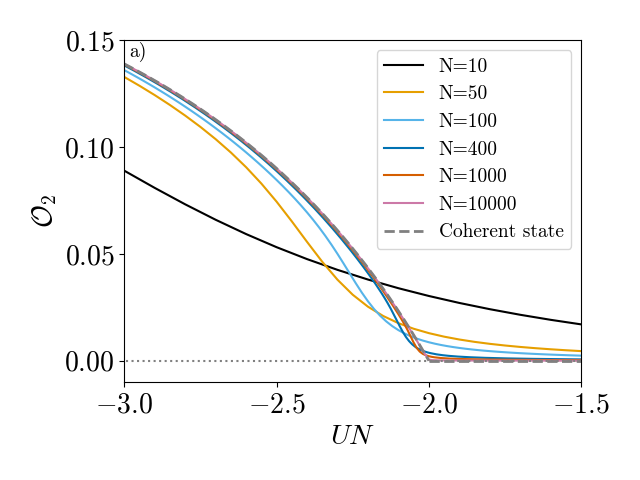}\\
    \centering\includegraphics[width=0.9\linewidth]{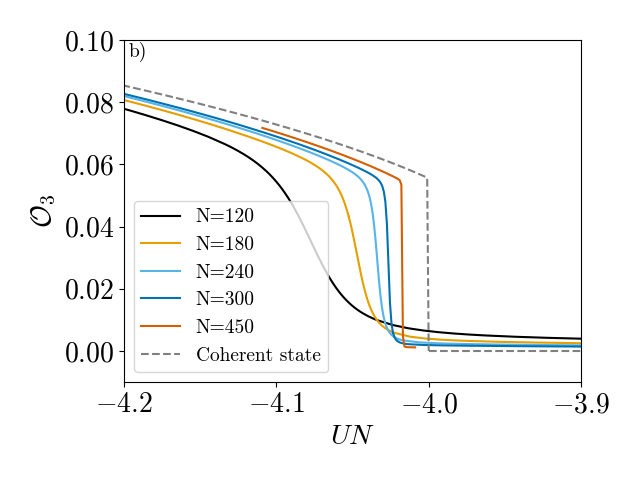}\\
    \centering\includegraphics[width=0.9\linewidth]{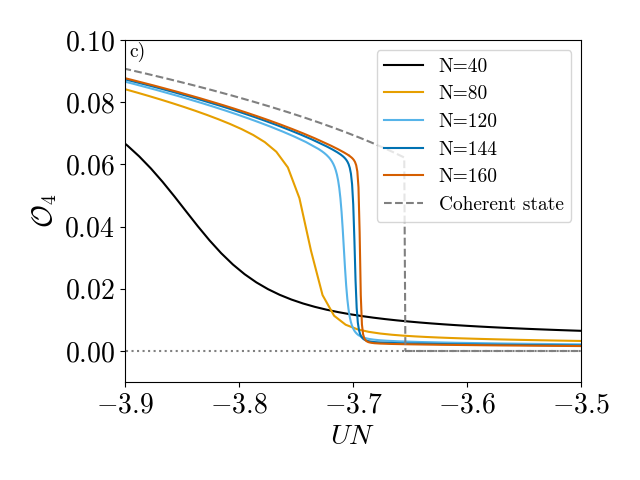}
    \caption{Order parameter $\mathcal{O}_L$ (see Fig. \ref{fig:OP0T}) shown in the vicinity of the phase transition, occurring at $NU_c=-2$ for $L=2$, $NU_c \simeq -4.00$ for $L=3$, and $NU_c \simeq -3.65$ for $L=4$. Despite sizeable finite-$N$ effects for $L > 2$, the qualitative difference between the second-order transition at $L=2$ and the first-order transitions observed for $L=3$ and $L=4$ is clearly visible.}
    \label{fig:OP0T-zoom}
\end{figure}

Importantly, a zoom-in view of these figures close to $U_c$, appearing on Fig. \ref{fig:OP0T-zoom}, reveals a difference between the $L=2$ and the $L>2$ QPT: while the former is second-order, the latter are first order,
as shown for $L=3$ and $4$ in Fig. \ref{fig:OP0T-zoom}. It corresponds to a discontinuity of $\{ \mu_j^m, j=1 \cdots L \}$ while tuning $U$ across $U_c$, as already discussed in section Sec. \ref{large_N_minim}.
Finally, it is useful to point out that an Hartree-Fock approach (or mean-field decoupling) of the SU(N) Hamiltonian (cf Eq. \eqref{eq:Ham}), similar to what has been done for other SU(N) systems (with both fermions and spins, on different lattices and with different fillings \cite{Rokhsar_1990,paramekanti_2007,hermele_topological_2011,Dong_2018,Feng_2023,Huang_2023,Zhang_2025}) leads to the very same results than the previous HP calculations at leading order in N, for both the critical interaction $U_c$ and  the order of parameters, as shown in Appendix Section \ref{app:MF}.

\section{Finite temperature transitions}
\label{finite_temp}
We consider finite-temperature transitions for $M=N$ particles.
The partition function of the system is:
\begin{equation}
    Z_{L, N} = \sum\limits_{\mathrm{irrep} \, \alpha} D^N_{\alpha} \text{Tr} e^{-\beta H_{\overline{\alpha}}},
\end{equation}
where the summation runs over all the $N$-box irreps $\alpha$ of SU($N$), $D_\alpha^N$ is the multiplicity (resp. dimension) of the $U(L)$ (resp. $SU(N)$) irrep $\overline{\alpha}$ (resp. $\alpha$), while $H_{\overline{\alpha}}$ is the effective Hamiltonian in the $U(L)$ orbital irrep $\overline{\alpha}$ of dimension $D_{\overline{\alpha}}^L$.\\

For positive $U$, we do not expect a finite-temperature transition towards the infinite temperature limit.
On the other hand, for $U \leq U_c < 0$, the ferromagnetic phase is distinct from the infinite-temperature paramagnet.
The Mermin-Wagner theorem does not prevent such a scenario: the chosen thermodynamic limit ($N\rightarrow +\infty$, $L$ constant) is an infinite-dimensional limit, explaining also the success of the mean-field description at $T = 0$ despite the strong interactions.\\

We start with the simplest case: $L=2$.

\subsection{Interplay between energies and multipliticities in the Boltzmann weights for $L=2$} \label{L2Tnonzero}
At finite-temperature, the exact mapping between the LMG model and the SU(N) 2-sites FHM (implying energy spectrum equivalence for a given N-box  irrep $\overline{\alpha}$ \cite{Nataf_2025}) no longer holds, because the associated multiplicities differ.
%
%
For the orbital irrep $\overline{\alpha}=(M_1,M_2)$, (where $N=M_1+M_2$), the multiplicity in the LMG model is the number of standard Young tableaux \cite{nataf2014} of shapes $\overline{\alpha}$ (or $\alpha$), that is to say:
\begin{equation}
    D_{\alpha}^\mathrm{LMG} = \binom{N}{M_1} \frac{M_1 + 1 - M_2}{M_1 + 1}.
\end{equation}
On the other hand, the multiplicity for the SU($N$) model is the dimension of $\alpha$ seen as an irrep of SU($N$) \cite{Nataf_2025}:
\begin{equation}
    D^N_{\alpha} = \binom{N}{M_1}\binom{N}{M_2} \frac{(N+1)(M_1-M_2+1)}{(M_1+1)(N-M_2+1)}. \label{binomial_D_SUN}
\end{equation}

From a methodological standpoint, although the phase diagram of the LMG model is accessible through simple mean-field calculation \cite{Pearce_1975,Blin_1996,Das_2006,Wilms_2012}, the mismatch in multiplicities  prevents the application of mean-field methods to the SU(N) FHM. Fortunately, a large-N expansion based on a saddle-point approximation of the partition function can be carried out to determine analytically the phase diagram of the SU(N) FHM for $L=2$, as shown below.

\begin{figure}[ht!]
    \centering\includegraphics[width=0.9\linewidth]{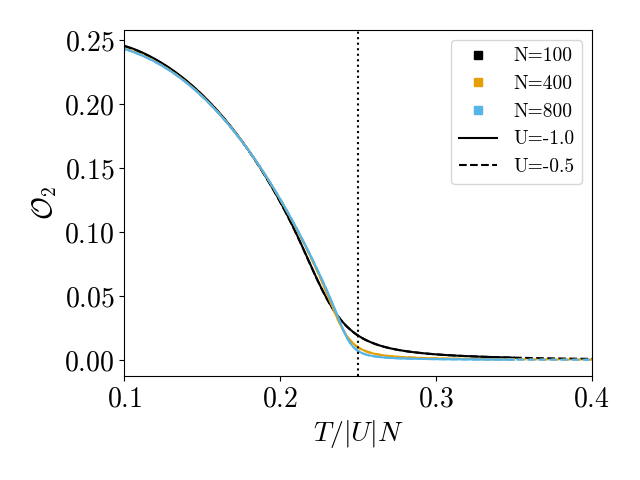}\\
    \centering\includegraphics[width=0.9\linewidth]{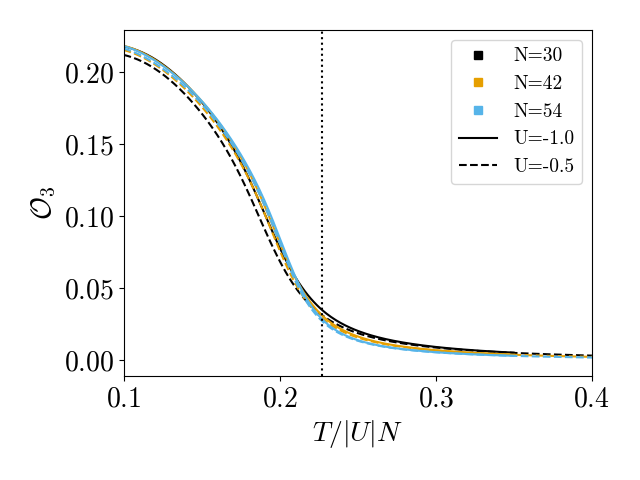}\\
    \centering\includegraphics[width=0.9\linewidth]{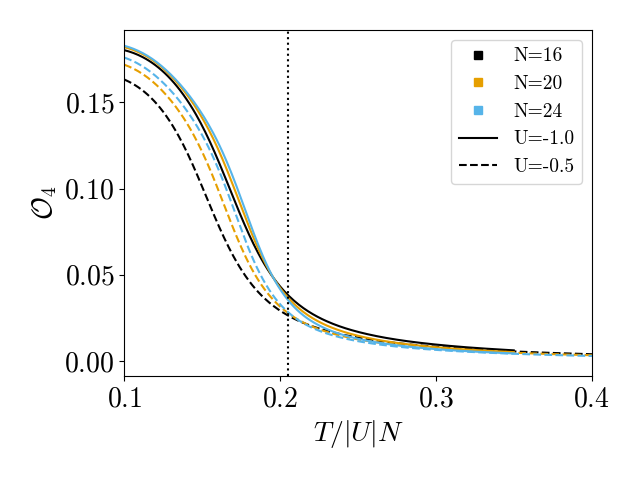}
    \caption{Order parameter $\mathcal{O}_L$ in the thermal state of the $L=2$ (top),  $L=3$ (middle) and $L=4$ (bottom) SU(N) model at filling $M = N$, for several values of $U$. For the larger $N$, we observe the predicted collapse as a function of $\beta U N$ only. The vertical line denotes the predicted finite-temperature transition (cf Eq. \eqref{betacL2}, \eqref{betacL3}, \eqref{betacL4} ), and agrees with the numerical data.}
    \label{fig:OPT}
\end{figure}

Firstly, for $\overline{\alpha}=(M_1,M_2)$, with $M_1+M_2=M=N$ and pseudo-spin $S=N/2-M_2$, the spectrum for $L=2$ sites can be approximated by the spectrum of an harmonic oscillator whose energies are ($k \in \mathbf N$):
\begin{equation}
\mathcal{E}_k=E_{G}+k \omega_b,
\end{equation}
where the ground state energy $E_{G}$ and the eigenfrequency $\omega_b$ read in the broken symmetry phase (i.e. for $U \leq U_c$ at $t=1$) \cite{Dusuel_2005,Das_2006,Campbell_2015,Pal_2023}:
\begin{align}
E_{G}&=U\{S^2+S(1-\sqrt{1-(1/SU)^2})+\frac{1}{U^2}+\frac{M^2}{4}\}, \\
\omega_b&=2\sqrt{(SU)^2-1}.
\end{align}
Note that the spectrum is twice degenerate.\\

Introducing $0 \leq \varepsilon_2 = M_2/N \leq 1/2$ and using the Stirling formula for the large-N expansion of the binomials in Eq. \eqref{binomial_D_SUN}, one can approximate $Z_{2, N}$ as:
\begin{equation}
    Z_{2, N} \approx \int\limits d\varepsilon_2 e^{- \beta \frac{UN^2}{4}(1-2\varepsilon_2)^2 + 2N \psi(\varepsilon_2) + o(N)},
\end{equation}
where the entropy function
\begin{equation}
 \psi(\varepsilon) = -\varepsilon\log  \varepsilon -  (1 - \varepsilon) \log  (1 - \varepsilon)
\end{equation}
is an approximation of $\frac{1}{N} \log \binom{N}{N\varepsilon}$ \cite{Wilms_2012}.
At  fixed $\beta$ in $Z_{2, N}$, there is a trade-off to find between the Boltzmann weights $e^{-\beta \mathcal{E}_k}$ 
and the multiplicities $D_{\alpha}^N$ to determine the most relevant irrep $\overline{\alpha}$ of pseudo-spin spin $S=N/2(1-2\varepsilon_2)$.

The saddle-point approximation in the large-N limit of $Z_{2, N}$ requires then the maximization of:
\begin{equation}
    -\beta UN(1-2\varepsilon_2)^2 + 8\psi(\varepsilon_2). \label{to_max}
\end{equation}
Let's  define $\tilde{\beta} = \beta  \vert U\vert  N=-\beta  U  N \geq 0$.
For $\tilde{\beta} =0$, the irrep with the largest $\psi(\varepsilon_2)$ will dominate, corresponding to $\varepsilon_2 =\varepsilon_2^*= 1/2$: this defines the infinite temperature phase.
%
%
%

For $\tilde{\beta} >0$, the extrema $\varepsilon_2^*<1/2$ satisfy
\begin{equation}
    \tilde{\beta} = \frac{1}{(\varepsilon_2^* - 1/2)}\log \frac{\varepsilon_2^*}{1-\varepsilon_2^*}.\label{eq:betaFTL2}
\end{equation}

\begin{figure}
    \centering
    \includegraphics[width=0.9\linewidth]{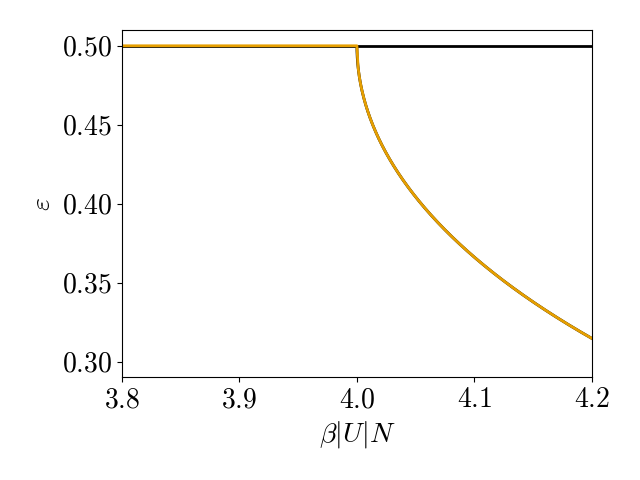}
  \includegraphics[width=0.9\linewidth]{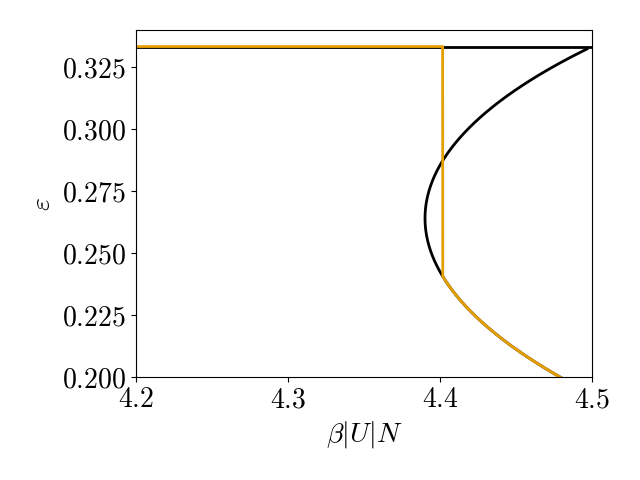}
    \caption{Top: solutions of Eq.~\eqref{eq:betaFTL2} for $L=2$ sites. The orange line single out the solution minimizing the free energy. 
    A continuous second-order phase-transition occurs at $\beta_c \vert U \vert N  = 4$.
    Bottom: solutions of Eq.~\eqref{eq:betaFTL3} for L=3. The orange line singles out the minimum of the free energy. The discontinuity at $\beta_c \vert U \vert N  \approx 4.40$ is a signature
    of a first-order phase transition.
    }
    \label{fig:FTL2}
\end{figure}

The RHS in Eq.~\eqref{eq:betaFTL2} is strictly decreasing with $\varepsilon_2^*$, and always correspond to a maximum of Eq.~\eqref{to_max}, with $\tilde{\beta}(\frac{1}{2}^-) = 4$.
Consequently, we expect a phase transition at $\tilde{\beta} = \tilde{\beta}_c=4$: 

For $\tilde{\beta} <\tilde{\beta}_c$, $\varepsilon_2^*=1/2 \Rightarrow M_1=M_2=N/2$, the configurations that dominate are  such that each site is equally populated with $N/2$ fermions (cf Fig. \ref{fig:sketch}, b) ).
The order of parameter $\mathcal{O}_2$ is also close to 0, as shown in Fig. \ref{fig:OPT} a).

\_For $\tilde{\beta} >\tilde{\beta}_c$, an other maximum appears for an intermediate irrep $\overline{\alpha}$, i.e. with $0 \leq \varepsilon_2^*<1/2$: the most likely configurations satisfy $M_1>M_2$, $\mathcal{O}_2$ increases with increasing $\tilde{\beta}$ (or decreasing $T$) as shown in Fig. \ref{fig:OPT} a).
 Moreover, the {\bf continuous} evolution of $\varepsilon_2^*$ with $\beta  \vert U\vert  N$ across the transition, shown in Fig. \ref{fig:FTL2} top, indicates a second order phase transition.

\begin{figure}[ht!]
\centering\includegraphics[width=0.9\linewidth]{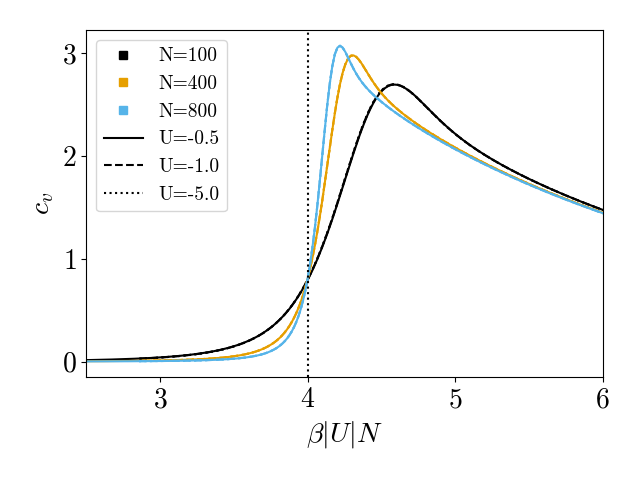}\\
       \centering\includegraphics[width=0.9\linewidth]{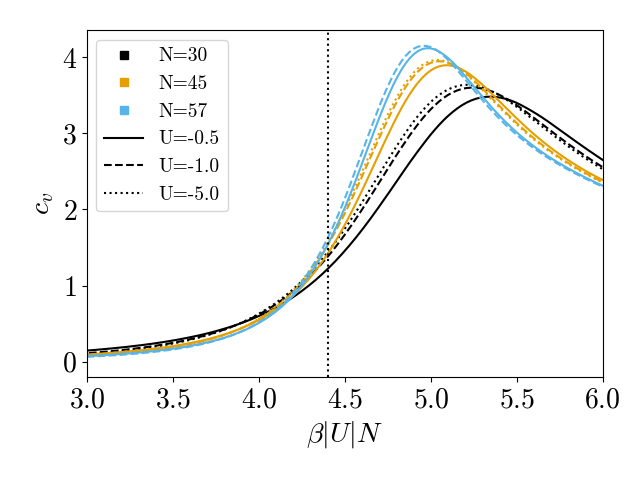}\\
    \centering\includegraphics[width=0.9\linewidth]{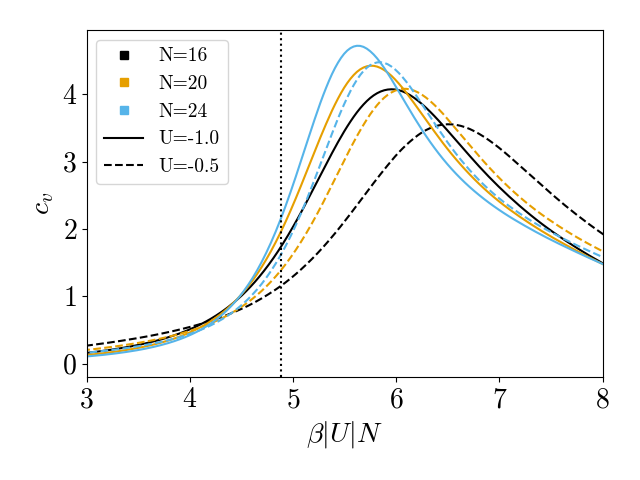}
    \caption{Specific heat $c_v$ defined as $\frac{dE}{NdT}$ of the $L=2$ (top), $L=3$ (middle) and $L=4$ (bottom) SU(N) model at filling $M = N$, for several values of $U$. The transition is well-marked by the drifting peak of the specific heat converging towards L-dependent values of $-\beta_c U N$ (cf Eq. \eqref{betacL2}, \eqref{betacL3}, \eqref{betacL4}) shown as vertical dotted lines. For $L > 2$, fully diagonalizing all symmetry sectors strongly limit our computation.}
        \label{fig:CVT}
\end{figure}

Finally, the large-N phase diagram for $L=2$ is also confirmed by the behavior of the specific heat per particle $c_v= \frac{1}{N} dE/dT$ at finite N, as shown in Fig.~\ref{fig:CVT}, where the location of the maximum of $c_v$ converges towards $\tilde{\beta}=4$. 
%

\subsection{$L \geq 3$}
The protocole developed for $L = 2$ in Sec. \ref{L2Tnonzero} can be extended to $L\geq 3$ as follows:
For  $L = 3$, the multiplicity of an irrep  $\overline{\alpha}=[M_1, M_2, M_3]$ (where $N=M=M_1+M_2+M_3$) is

\begin{multline}
    D^N_{\alpha} =  \frac{\binom{N}{M_1}\binom{N}{M_2}\binom{N}{M_3} (N+1)^2(N+2)}{(M_1 + 1)(M_1 + 2)(N - M_2 + 1)} \\
    \frac{(M_1 - M_2 + 1) (M_2 - M_3 + 1)(M_1 - M_3 + 2)}{(M_2 + 1) (N - M_3 + 1) (N- M_3 + 2)}
\end{multline}
In the symmetry-broken phase, for large $\vert U \vert N$, the groundstate energy is approximated by the interaction energy.
For a given YD $\overline{\alpha}=[M_1, M_2, M_3]$, maximizing the occupation on site $1$, then on site $2$, then on site $3$, leads \footnote{One can think of the constraints associated with SSYT.} to an energy of the form:
\begin{equation}
    E(M_2, M_3) = \frac{U}{2} \left( (N - M_2 - M_3)^2 + M_2^2 + M_3^2\right).
\end{equation}
Introducing $\varepsilon_j = M_j/N$ (with the constraints that $0\leq \varepsilon_2 \leq 1/2$, $0\leq \varepsilon_3\leq \min(1/3, \varepsilon_2, 1 - 2\varepsilon_2)$), the saddle point approximation requires to maximize the function
\begin{multline}
    \frac{N\tilde{\beta}}{2} ((1 - \varepsilon_2 -\varepsilon_3)^2 + \varepsilon_2^2+\varepsilon_3^2) \\
    + N(\psi(\varepsilon_2) + \psi(\varepsilon_3) + \psi(\varepsilon_2 + \varepsilon_3)).
\end{multline}
The extrema verify the system of coupled equations
\begin{align}
    \tilde{\beta} (2 \varepsilon_2 + \varepsilon_3 -1) &= \log \frac{\varepsilon_2}{(1 - \varepsilon_2)} + \log \frac{\varepsilon_2 + \varepsilon_3}{(1 - \varepsilon_2 - \varepsilon_3)}\label{eq:FTL3} \\ 
    \tilde{\beta} (2 \varepsilon_3 + \varepsilon_2 -1) &= \log \frac{\varepsilon_3}{(1 - \varepsilon_3)} + \log \frac{\varepsilon_2 + \varepsilon_3}{(1 - \varepsilon_2 - \varepsilon_3)}.
\end{align}
Like before, the infinite-temperature limit  ($\varepsilon_2 = \varepsilon_3 = 1/3$) is always an extrema.
Numerically, we find that the system only admits solutions for $\varepsilon_2 = \varepsilon_3=\varepsilon$, leading to 
\begin{equation}
    \tilde{\beta}= \frac{1}{3\varepsilon-1} \left(\log \frac{\varepsilon}{1 - \varepsilon} + \log \frac{2\varepsilon}{1 - 2\varepsilon} \right).\label{eq:betaFTL3}
\end{equation}
Contrarily to $L = 2$, $\tilde{\beta}$ is not a monotonic function of $\varepsilon$ anymore, but admits a minimum at $\varepsilon^* \approx 0.264$ for $\tilde{\beta}_{3}^* \approx 4.3902$.
As in the zero temperature limit, requiring $\tilde{\beta}$ to be a (global) minimum of the free energy pushes back the transition to $\tilde{\beta}_{3, c} \approx 4.4019$ with $\varepsilon_{c} \approx 0.2407$.

The sharp jump in $\varepsilon$ at $\tilde{\beta}_{3, c} $ is symptomatic of a first-order phase transition, as exemplified in the bottom of Fig.~\ref{fig:FTL2} .
The large-N critical temperature agrees with our exact numerical simulations, though we cannot distinguish whether $\beta_{3, c}$ or $\beta_3^*$ is the correct critical point.
Indeed, finite-size effects are significant, and the first-order nature of the transition is not visible within the order parameter (cf Fig. \ref{fig:OPT})

\begin{figure}
    \centering
    \includegraphics[width=1\linewidth]{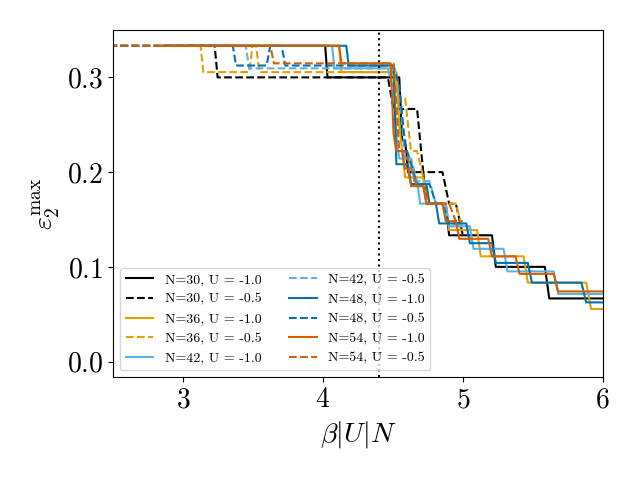}
    \includegraphics[width=1\linewidth]{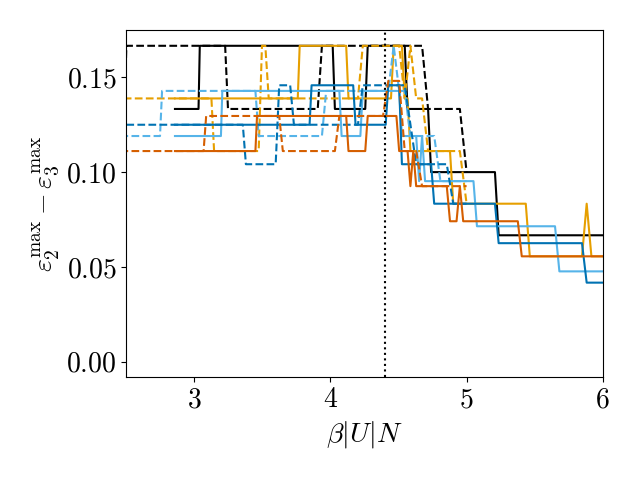}
    \caption{Top: $\varepsilon_2$ minimizing the free energy for $L=3$. It jumps at the critical temperature, in agreement with the saddle-point prediction. Bottom: $\varepsilon_2-\varepsilon_3$ for the minimizing irrep. Contrarily to the predicted thermodynamic limit, we see that $\varepsilon_2-\varepsilon_3$ remains significantly larger than $0$, underlining the significant finite-size effects. Vertical dashed lines represent Eq. \eqref{betacL3}}
    \label{fig:FiniteT-L3-analytic}
\end{figure}

Instead, we determine the irrep (characterized by the couple $(\varepsilon_2, \varepsilon_3)$) that numerically minimizes the free energy. 
As predicted, a stable discontinuity close to the predicted critical temperature is evidenced in Fig. \ref{fig:FiniteT-L3-analytic}, in agreement with the first-order scenario.
Nonetheless, the finite-size effects are significant: in the thermodynamic limit, we expect $\varepsilon_2 = \varepsilon_3$ in all minimizing irreps, but
even at $N > 50$, we instead observe in Fig. \ref{fig:FiniteT-L3-analytic} a significant residual $\varepsilon_2 - \varepsilon_3$.\\

\paragraph{$L \geq 3$}
 We can generalize this approach to any $L$.
The dominant term in the free energy reads
\begin{equation}
    \frac{\tilde{\beta}N}{2} \sum\limits_j \varepsilon_j^2 +  N \sum\limits_j \psi(\varepsilon_j),
\end{equation}
with $\varepsilon_1 = 1 - \sum\limits_{j >1} \varepsilon_j$.

 Looking again for symmetric solutions, we obtain the equation
\begin{equation}
    \tilde{\beta}= \frac{1}{L\varepsilon - 1} \left( \log \frac{\varepsilon}{1-\varepsilon} + \log \frac{(L-1)\varepsilon}{1-(L-1)\varepsilon} \right).
\end{equation}
In particular, for $L = 4$, we predict a first-order phase transition for $\beta_{4, c} \vert U \vert  N \approx 4.878$ and $\varepsilon_c \approx 0.134$, with again appearance of other extremas at $\beta_{4}^* \vert U \vert N \approx 4.828$ and $\varepsilon^* \approx 0.164$. The finite-temperature order of parameters $\mathcal{O}_4$ plotted in Fig. \ref{fig:OPT},
as well as the specific heat shown in Fig. \ref{fig:CVT} are compatible with this large-N prediction.
To summarize, the large-N results for finite-temperature transitions involve a critical temperature $T_c=1/\beta_c$ that satisfies:
\begin{align}
\beta_cUN &=-4,  \hspace{2.4cm} \text{for} \hspace{.4 cm} L=2, \label{betacL2}\\
\beta_cUN &= -(4.40\pm 0.01)   \hspace{0.8cm} \text{for} \hspace{0.5 cm} L=3,\label{betacL3}\\
\beta_cUN &=-(4.85\pm 0.05)  \hspace{0.8cm} \text{for} \hspace{0.5 cm} L=4.\label{betacL4}
\end{align}
The linear dependence of $T_c$ with both $-U$ and $N$ has the following consequence: at fixed temperature $T$ and number of sites $L$, a {\it larger} number of colors $N$ means a {\it smaller} (but still attractive) $\vert U \vert$ to break the $\mathbb{Z}_L$ symmetry and {\it condense} all the atoms on a given site.

Contrarily to zero temperature, we are strongly limited in system sizes $L$ as we need to fully diagonalize the Hamiltonian over all sectors, but we expect such a behavior to hold for $L>4$.

\section{Conclusions and Perspectives}
To conclude, we have extended the results of \cite{Nataf_2025} to both $L>2$ and $T>0$, using analytical large-N treatment supported by ED with full SU(N) symmetry.
While the analytical methods (L-levels HP transformation, saddle point approximation) are rather standard, their use is, to the best of our knowledge, original to study the SU(N) FHM on a L-sites ring for a large number of colors $N$ (equal to the number of particles) and a small number of sites $L$.

They lead to simple phase diagrams with measurables predictions: the order of parameters  
$\mathcal{O}_L$, which is the number of on-site pair per site, could be accessed through photoassociation  process \cite{takahashi2012,taie2020observation,Fallani_2022,Pasqualetti_2024}, and would allow to distinguish between the second order transition for $L=2$, and the first order transition for $L>2$, at predicted values of the critical interaction amplitude $U_c$ (cf Tab. \ref{tab:placeholder}).
Regarding the critical temperature $T_c$, which depends linearly on both $N$ and $-U>0$, the onset of the $\mathbb{Z}_L$ symmetry-breaking  condensation requires a smaller magnitude of the local attractive interaction $\vert U \vert$ as $N$ increases at fixed temperature, another theoretical outcome of our work.

About the experimental realization of attractive interaction, interorbital Feshbach resonances have already been shown to be suitable for this purpose
in the case of $^{\text{173}}$Yt \cite{Pagano_2015,Hofer_2015}, while the proposal of using ultracold molecules controlled by external electric fields appears viable
especially for large-N fermions \cite{Mukherjee_2024}.

Finally, in view of the increasing theoretical interest in SU(N) Fermi-Hubbard systems with attractive interactions \cite{Rapp_2007,Inaba_2009,Titvinidze_2011,Pohlmann_2013,Chetcuti_2023,Chetcuti_2023b,Xu_2023,Li_2025,Stepp_2025}, our results contribute to the thermodynamic characterization of the attractive SU(N) Fermi-Hubbard model,
which has been less studied than the thermodynamics of repulsive SU(N) Fermi-Hubbard model\cite{hazzard,Yanatori_2016,Ibarra_2021,He_2025}.
In this respect, it opens multiple perspectives in several directions. 
Since the $L$-fold degeneracy in the $U/t \rightarrow -\infty$ originates from the breaking of the translation symmetry operator, one could analyze how our results extend when the boundary conditions or more generally the lattice parameters (such as e.g. the dimension, the hopping amplitudes, the range of the atom-atom interaction, etc...) change.
In particular, modulation of the hopping amplitudes, could break {\it partially} the lattice symmetries \cite{Capponi_2025}, changing the finite temperature phase diagram.
An other intriguing question is the following: for the same filling (i.e. $M=N$), what are the minimal ingredients to force the ($T=0$) ground state not to live in the fully symmetric irrep?
Adding more orbitals \cite{Kobayashi_2012,Bois_2015,Capponi_annals_2016}, could be a way.
In any case, implementing additional sites or more (non-degenerate) orbitals may require methodological adjustments, but the color factorization scheme \cite{Botzung_2023_PRL,Botzung_2023}  should be used, as it significantly enhances the effectiveness of numerical methods for SU(N)-invariant systems.

{\it Acknowledgments}
P.N. is supported by the IRP EXQMS project from CNRS. L.H. acknowledges the Tremplin funding from CNRS Physique and was also supported by the ANR JCJC ANR-25-CE30-2205-01. 

This work is dedicated to the memory of Bart van Tiggelen.

\bibliographystyle{unsrt}
\bibliography{bibliography}

\begin{thebibliography}{10}

\bibitem{Hubbard_1963}
J.~Hubbard.
\newblock {Electron correlations in narrow energy bands}.
\newblock {\em Proc. R. Soc. London A - Math. Phys. Sci.}, 276(1365):238--257,
  November 1963.

\bibitem{Gutzwiller_1963}
Martin~C. Gutzwiller.
\newblock Effect of correlation on the ferromagnetism of transition metals.
\newblock {\em Phys. Rev. Lett.}, 10:159--162, Mar 1963.

\bibitem{Scalapino_2012}
D.~J. Scalapino.
\newblock {A common thread: The pairing interaction for unconventional
  superconductors}.
\newblock {\em Rev. Mod. Phys.}, 84(4):1383--1417, October 2012.

\bibitem{review_Arovas_2022}
Daniel~P. Arovas, Erez Berg, Steven~A. Kivelson, and Srinivas Raghu.
\newblock The hubbard model.
\newblock {\em Annual Review of Condensed Matter Physics}, 13(1):239--274,
  2022.

\bibitem{review_Corboz_2022}
Mingpu Qin, Thomas Sch\"{a}fer, Sabine Andergassen, Philippe Corboz, and
  Emanuel Gull.
\newblock The hubbard model: A computational perspective.
\newblock {\em Annual Review of Condensed Matter Physics}, 13(1):275--302,
  2022.

\bibitem{Anderson_1987}
P.~W. Anderson.
\newblock The resonating valence bond state in la<sub>2</sub>cuo<sub>4</sub>
  and superconductivity.
\newblock {\em Science}, 235(4793):1196--1198, 1987.

\bibitem{Rice_1988}
F.~C. Zhang and T.~M. Rice.
\newblock Effective hamiltonian for the superconducting cu oxides.
\newblock {\em Phys. Rev. B}, 37:3759--3761, Mar 1988.

\bibitem{Assaraf_1999}
Roland Assaraf, Patrick Azaria, Michel Caffarel, and Philippe Lecheminant.
\newblock Metal-insulator transition in the one-dimensional $\mathrm{SU}(n)$
  hubbard model.
\newblock {\em Phys. Rev. B}, 60:2299--2318, Jul 1999.

\bibitem{wu_exact_2003}
Congjun Wu, Jiang-ping Hu, and Shou-cheng Zhang.
\newblock Exact so(5) symmetry in the spin-$3/2$ fermionic system.
\newblock {\em Phys. Rev. Lett.}, 91:186402, Oct 2003.

\bibitem{Honerkamp2004}
Carsten Honerkamp and Walter Hofstetter.
\newblock Ultracold fermions and the $\mathrm{SU}(n)$ hubbard model.
\newblock {\em Phys. Rev. Lett.}, 92:170403, Apr 2004.

\bibitem{Capponi_annals_2016}
S.~Capponi, P.~Lecheminant, and K.~Totsuka.
\newblock Phases of one-dimensional su(n) cold atomic fermi gases from
  molecular luttinger liquids to topological phases.
\newblock {\em Annals of Physics}, 367:50 -- 95, 2016.

\bibitem{Ibarra_Garcia_Padilla_2024}
Eduardo Ibarra-Garcia-Padilla and Sayan Choudhury.
\newblock Many-body physics of ultracold alkaline-earth atoms with
  su(n)-symmetric interactions.
\newblock {\em Journal of Physics: Condensed Matter}, 37(8):083003, dec 2024.

\bibitem{Chen_2024}
Gang~V. Chen and Congjun Wu.
\newblock Multiflavor mott insulators in quantum materials and ultracold atoms.
\newblock {\em npj Quantum Materials}, 9(1):1, 2024.

\bibitem{affleck_exact_1986}
Ian Affleck.
\newblock Exact critical exponents for quantum spin chains, non-linear
  $\sigma$-models at $\theta=\pi$ and the quantum hall effect.
\newblock {\em Nuclear Physics B}, 265(3):409 -- 447, 1986.

\bibitem{Affleck_1988}
I~Affleck.
\newblock {Critical behaviour of SU(n) quantum chains and topological
  non-linear $\sigma$-models}.
\newblock {\em Nuclear Physics B}, 305:582--596, 1988.

\bibitem{Rokhsar_1990}
Daniel~S. Rokhsar.
\newblock Quadratic quantum antiferromagnets in the fermionic large-n limit.
\newblock {\em Phys. Rev. B}, 42:2526--2531, Aug 1990.

\bibitem{Marder_1990}
M.~Marder, N.~Papanicolaou, and G.~C. Psaltakis.
\newblock Phase separation in a t-j model.
\newblock {\em Phys. Rev. B}, 41:6920--6932, Apr 1990.

\bibitem{Read_1991}
N.~Read and Subir Sachdev.
\newblock Large-n expansion for frustrated quantum antiferromagnets.
\newblock {\em Phys. Rev. Lett.}, 66:1773--1776, Apr 1991.

\bibitem{Chung_2001}
C~H Chung, J~B Marston, and Ross~H McKenzie.
\newblock Large-n solutions of the heisenberg and hubbard-heisenberg models on
  the anisotropic triangular lattice: application to cs2cucl4 and to the
  layered organic superconductors $\kappa$-(bedt-ttf)$_2$x
  (bedt-ttf$\equiv$bis(ethylene-dithio)tetrathiafulvalene); x$\equiv$anion).
\newblock {\em Journal of Physics: Condensed Matter}, 13(22):5159, jun 2001.

\bibitem{Veillette_2007}
Martin~Y. Veillette, Daniel~E. Sheehy, and Leo Radzihovsky.
\newblock Large-$n$ expansion for unitary superfluid fermi gases.
\newblock {\em Phys. Rev. A}, 75:043614, Apr 2007.

\bibitem{Polychronakos_2024}
Alexios~P. Polychronakos and Konstantinos Sfetsos.
\newblock Triple critical point and emerging temperature scales in su(n)
  ferromagnetism at large n.
\newblock {\em Nuclear Physics B}, 1009:116748, 2024.

\bibitem{Wu_review_2006}
Congjun Wu.
\newblock Hidden symmetry and quantum phases in spin-3/2 cold atomic systems.
\newblock {\em Modern Physics Letters B}, 20(27):1707--1738, 2006.

\bibitem{gorshkov_two_2010}
Alexey~Vyacheslavovich Gorshkov, M~Hermele, V~Gurarie, C~Xu, Paul~S Julienne,
  J~Ye, Peter Zoller, Eugene Demler, Mikhail~D Lukin, and AM~Rey.
\newblock Two-orbital su(n) magnetism with ultracold alkaline-earth atoms.
\newblock {\em Nature physics}, 6(4):289--295, 2010.

\bibitem{Cazalilla_2014}
Miguel~A Cazalilla and Ana~Maria Rey.
\newblock Ultracold gases of ytterbium: ferromagnetism and mott states in an
  su(6) fermi system.
\newblock {\em Rep. Prog. Phys.}, 77(12):124401, 2014.

\bibitem{takahashi2012}
S.~Taie, R.~Yamazaki, S.~Sugawa, and Y.~Takahashi.
\newblock {An SU(6) Mott insulator of an atomic Fermi gas realized by
  large-spin Pomeranchuk cooling}.
\newblock {\em Nat Phys}, 8(4):825--830, 2012.

\bibitem{Pagano2014}
Guido Pagano, Marco Mancini, Giacomo Cappellini, Pietro Lombardi, Florian
  Sch\"{a}fer, Hui Hu, Xia-Ji Liu, Jacopo Catani, Carlo Sias, Massimo Inguscio,
  and Leonardo Fallani.
\newblock {A one-dimensional liquid of fermions with tunable spin}.
\newblock {\em Nature Physics}, 10(3):198--201, February 2014.

\bibitem{Scazza2014}
F.~Scazza, C.~Hofrichter, M.~H\"{o}fer, P.~C. {De Groot}, I.~Bloch, and
  S.~F\"{o}lling.
\newblock Observation of two-orbital spin-exchange interactions with ultracold
  su(n)- symmetric fermions.
\newblock {\em Nature Physics}, 10(August):779--784, 2014.

\bibitem{Zhang2014}
X~Zhang, M~Bishof, S~L Bromley, C~V Kraus, M~S Safronova, P~Zoller, a~M Rey,
  and J~Ye.
\newblock {Spectroscopic observation of SU(N)-symmetric interactions in Sr
  orbital magnetism.}
\newblock {\em Science (New York, N.Y.)}, 345(m):1467--1473, August 2014.

\bibitem{Hofrichter_2016}
Christian Hofrichter, Luis Riegger, Francesco Scazza, Moritz H\"ofer, Diogo~Rio
  Fernandes, Immanuel Bloch, and Simon F\"olling.
\newblock Direct probing of the mott crossover in the $\mathrm{SU}(n)$
  fermi-hubbard model.
\newblock {\em Phys. Rev. X}, 6:021030, Jun 2016.

\bibitem{Becker_2021}
B.~Abeln, K.~Sponselee, M.~Diem, N.~Pintul, K.~Sengstock, and C.~Becker.
\newblock Interorbital interactions in an
  $\mathrm{SU}(2)\ensuremath{\bigotimes}\mathrm{SU}(6)$-symmetric fermi-fermi
  mixture.
\newblock {\em Phys. Rev. A}, 103:033315, Mar 2021.

\bibitem{taie2020observation}
Shintaro Taie, Eduardo Ibarra-Garc{\'\i}a-Padilla, Naoki Nishizawa, Yosuke
  Takasu, Yoshihito Kuno, Hao-Tian Wei, Richard~T. Scalettar, Kaden R.~A.
  Hazzard, and Yoshiro Takahashi.
\newblock Observation of antiferromagnetic correlations in an ultracold su(n)
  hubbard model.
\newblock {\em Nature Physics}, 18(11):1356--1361, 2022.

\bibitem{Fallani_2022}
D.~Tusi, L.~Franchi, L.~F. Livi, K.~Baumann, D.~Benedicto~Orenes, L.~Del~Re,
  R.~E. Barfknecht, T.~W. Zhou, M.~Inguscio, G.~Cappellini, M.~Capone,
  J.~Catani, and L.~Fallani.
\newblock Flavour-selective localization in interacting lattice fermions.
\newblock {\em Nature Physics}, 18(10):1201--1205, 2022.

\bibitem{Pasqualetti_2024}
G.~Pasqualetti, O.~Bettermann, N.~Darkwah~Oppong,
  E.~Ibarra-Garc\'{\i}a-Padilla, S.~Dasgupta, R.~T. Scalettar, K.~R.~A.
  Hazzard, I.~Bloch, and S.~F\"olling.
\newblock Equation of state and thermometry of the 2d $\mathrm{SU}(n)$
  fermi-hubbard model.
\newblock {\em Phys. Rev. Lett.}, 132:083401, Feb 2024.

\bibitem{Mukherjee_2024}
Bijit Mukherjee, Jeremy~M Hutson, and Kaden Hazzard.
\newblock Su(n) magnetism with ultracold molecules.
\newblock {\em New Journal of Physics}, 2024.

\bibitem{Salfi_2016}
J.~Salfi, J.~A. Mol, R.~Rahman, G.~Klimeck, M.~Y. Simmons, L.~C.~L. Hollenberg,
  and S.~Rogge.
\newblock Quantum simulation of the hubbard model with dopant atoms in silicon.
\newblock {\em Nature Communications}, 7(1):11342, 2016.

\bibitem{Wang_2022}
Xiqiao Wang, Ehsan Khatami, Fan Fei, Jonathan Wyrick, Pradeep Namboodiri,
  Ranjit Kashid, Albert~F. Rigosi, Garnett Bryant, and Richard Silver.
\newblock Experimental realization of an extended fermi-hubbard model using a
  2d lattice of dopant-based quantum dots.
\newblock {\em Nature Communications}, 13(1), 11 2022.

\bibitem{Wei_2024}
Hao-Tian Wei, Eduardo Ibarra-Garc\'{\i}a-Padilla, Michael~L. Wall, and Kaden
  R.~A. Hazzard.
\newblock Hubbard parameters for programmable tweezer arrays.
\newblock {\em Phys. Rev. A}, 109:013318, Jan 2024.

\bibitem{Chew_2024}
Y.~T. Chew, M.~Poitrinal, T.~Tomita, S.~Kitade, J.~Mauricio, K.~Ohmori, and
  S.~de~L\'es\'eleuc.
\newblock Ultraprecise holographic optical tweezer array.
\newblock {\em Phys. Rev. A}, 110:053518, Nov 2024.

\bibitem{Nataf_2025}
Pierre Nataf.
\newblock Su($n$) fermi-hubbard model on two sites: Bethe ansatz solution and
  quantum phase transition of the lipkin-meshkov-glick model in the large-$n$
  limit.
\newblock {\em Phys. Rev. A}, 111:L020201, Feb 2025.

\bibitem{LMG_1965}
H.J. Lipkin, N.~Meshkov, and A.J. Glick.
\newblock Validity of many-body approximation methods for a solvable model:
  (i). exact solutions and perturbation theory.
\newblock {\em Nuclear Physics}, 62(2):188--198, 1965.

\bibitem{Gelfand_1950}
I.~M. Gelfand and M.~L. Tsetlin.
\newblock Finite dimensional representations of the group of unimodular
  matrices.
\newblock {\em Dokl. Akad. Nauk SSSR}, 71:825, 1950.

\bibitem{Botzung_2023_PRL}
Thomas Botzung and Pierre Nataf.
\newblock Exact diagonalization of $\mathrm{SU}(n)$ fermi-hubbard models.
\newblock {\em Phys. Rev. Lett.}, 132:153001, Apr 2024.

\bibitem{Botzung_2023}
Thomas Botzung and Pierre Nataf.
\newblock Numerical observation of su($n$) nagaoka ferromagnetism.
\newblock {\em Phys. Rev. B}, 109:235131, Jun 2024.

\bibitem{nataf2014}
Pierre Nataf and Fr\'ed\'eric Mila.
\newblock Exact diagonalization of heisenberg $\mathrm{SU}(n)$ models.
\newblock {\em Phys. Rev. Lett.}, 113:127204, Sep 2014.

\bibitem{itzykson}
C~Itzykson and M.~Nauenberg.
\newblock Unitary groups: Representations and decompositions.
\newblock {\em Rev. Mod. Phys.}, 38:95--120, Jan 1966.

\bibitem{Stellmer_2013}
Simon Stellmer, Rudolf Grimm, and Florian Schreck.
\newblock Production of quantum-degenerate strontium gases.
\newblock {\em Phys. Rev. A}, 87:013611, Jan 2013.

\bibitem{Bataille_2020}
P.~Bataille, A.~Litvinov, I.~Manai, J.~Huckans, F.~Wiotte, A.~Kaladjian,
  O.~Gorceix, E.~Mar\'echal, B.~Laburthe-Tolra, and M.~Robert-de Saint-Vincent.
\newblock Adiabatic spin-dependent momentum transfer in an su($n$) degenerate
  fermi gas.
\newblock {\em Phys. Rev. A}, 102:013317, Jul 2020.

\bibitem{Ahmed_2025}
H.~Ahmed, A.~Litvinov, P.~Guesdon, E.~Mar\'echal, J.H. Huckans, B.~Pasquiou,
  B.~Laburthe-Tolra, and M.~Robert-de Saint-Vincent.
\newblock Coherent control over the high-dimensional space of the nuclear spin
  of alkaline-earth atoms.
\newblock {\em PRX Quantum}, 6:020352, Jun 2025.

\bibitem{Newman_1977}
C.~M. Newman and L.~S. Schulman.
\newblock {Metastability and the analytic continuation of eigenvalues}.
\newblock {\em Journal of Mathematical Physics}, 18(1):23--30, 01 1977.

\bibitem{HP_1940}
T.~Holstein and H.~Primakoff.
\newblock Field dependence of the intrinsic domain magnetization of a
  ferromagnet.
\newblock {\em Phys. Rev.}, 58:1098--1113, Dec 1940.

\bibitem{Papanicolaou_1984}
N.~Papanicolaou.
\newblock Pseudospin approach for planar ferromagnets.
\newblock {\em Nuclear Physics B}, 240(3):281--311, 1984.

\bibitem{Katriel_1983}
Jacob Katriel, Mario Rasetti, and Allan~I. Solomon.
\newblock Generalized holstein-primakoff squeezed states for su(n).
\newblock {\em Phys. Rev. D}, 35:2601--2602, Apr 1987.

\bibitem{papanicolaou1988}
N~Papanicolaou.
\newblock {Unusual Phases in Quantum Spin 1 Systems}.
\newblock {\em Nuclear Physics B}, 305:367--395, 1988.

\bibitem{Molmer_2010}
Z.~Kurucz and K.~M\o{}lmer.
\newblock Multilevel holstein-primakoff approximation and its application to
  atomic spin squeezing and ensemble quantum memories.
\newblock {\em Phys. Rev. A}, 81:032314, Mar 2010.

\bibitem{Romhanyi_2012}
Judit Romh\'anyi and Karlo Penc.
\newblock Multiboson spin-wave theory for ba${}_{2}$coge${}_{2}$o${}_{7}$: A
  spin-3/2 easy-plane n\'eel antiferromagnet with strong single-ion anisotropy.
\newblock {\em Phys. Rev. B}, 86:174428, Nov 2012.

\bibitem{joshi1999}
A.~Joshi, M.~Ma, F.~Mila, D.~N. Shi, and F.~C. Zhang.
\newblock Elementary excitations in magnetically ordered systems with orbital
  degeneracy.
\newblock {\em Phys. Rev. B}, 60:6584--6587, Sep 1999.

\bibitem{toth2010}
Tam\'as~A. T\'oth, Andreas~M. L\"auchli, Fr\'ed\'eric Mila, and Karlo Penc.
\newblock Three-sublattice ordering of the su(3) heisenberg model of
  three-flavor fermions on the square and cubic lattices.
\newblock {\em Phys. Rev. Lett.}, 105:265301, Dec 2010.

\bibitem{Kim_2017}
Francisco~H. Kim, Karlo Penc, Pierre Nataf, and Fr\'ed\'eric Mila.
\newblock Linear flavor-wave theory for fully antisymmetric su($n$) irreducible
  representations.
\newblock {\em Phys. Rev. B}, 96:205142, Nov 2017.

\bibitem{Hayn_2011}
Mathias Hayn, Clive Emary, and Tobias Brandes.
\newblock Phase transitions and dark-state physics in two-color superradiance.
\newblock {\em Phys. Rev. A}, 84:053856, Nov 2011.

\bibitem{Nataf_2012}
Pierre Nataf, Alexandre Baksic, and Cristiano Ciuti.
\newblock Double symmetry breaking and two-dimensional quantum phase diagram in
  spin-boson systems.
\newblock {\em Phys. Rev. A}, 86:013832, Jul 2012.

\bibitem{Baksic_2013}
Alexandre Baksic, Pierre Nataf, and Cristiano Ciuti.
\newblock Superradiant phase transitions with three-level systems.
\newblock {\em Phys. Rev. A}, 87:023813, Feb 2013.

\bibitem{Jiang_2025}
Hao Jiang, Ze-Yun Shi, Bo~Li, Long Wang, Xiu-Juan Dong, Ming Ma, Kai Chen,
  Cheng Chen, Cheng-Rui Wu, Dong-Yan LÃŒ, and Yuan Zhou.
\newblock Simulation of three-level dicke quantum phase transitions using
  solid-state spins synergistically coupled to acoustics and microwave.
\newblock {\em Physica Scripta}, 100(9):095111, sep 2025.

\bibitem{Gilmore_1979}
R.~Gilmore.
\newblock The classical limit of quantum nonspin systems.
\newblock {\em Journal of Mathematical Physics}, 20(5):891--893, 05 1979.

\bibitem{Ibarra_2021}
Eduardo Ibarra-Garc\'{\i}a-Padilla, Sohail Dasgupta, Hao-Tian Wei, Shintaro
  Taie, Yoshiro Takahashi, Richard~T. Scalettar, and Kaden R.~A. Hazzard.
\newblock Universal thermodynamics of an $\mathrm{SU}(n)$ fermi-hubbard model.
\newblock {\em Phys. Rev. A}, 104:043316, Oct 2021.

\bibitem{paramekanti_2007}
Arun Paramekanti and J~B Marston.
\newblock Su ( n ) quantum spin models: a variational wavefunction study.
\newblock {\em Journal of Physics: Condensed Matter}, 19(12):125215, 2007.

\bibitem{hermele_topological_2011}
M~Hermele and V~Gurarie.
\newblock Topological liquids and valence cluster states in two-dimensional
  {SU}(n) magnets.
\newblock {\em Physical Review B}, 84(17):1--24, November 2011.

\bibitem{Dong_2018}
Zhao-Yang Dong, Wei Wang, and Jian-Xin Li.
\newblock $\mathrm{SU}(n)$ spin-wave theory: Application to spin-orbital mott
  insulators.
\newblock {\em Phys. Rev. B}, 97:205106, May 2018.

\bibitem{Feng_2023}
Chunhan Feng, Eduardo Ibarra-Garc\'{\i}a-Padilla, Kaden R.~A. Hazzard, Richard
  Scalettar, Shiwei Zhang, and Ettore Vitali.
\newblock Metal-insulator transition and quantum magnetism in the su(3)
  fermi-hubbard model.
\newblock {\em Phys. Rev. Res.}, 5:043267, Dec 2023.

\bibitem{Huang_2023}
Chen-How Huang and Miguel~A Cazalilla.
\newblock Itinerant ferromagnetism in su(n)-symmetric fermi gases at finite
  temperature: first order phase transitions and time-reversal symmetry.
\newblock {\em New Journal of Physics}, 25(6):063005, jun 2023.

\bibitem{Zhang_2025}
Zewen Zhang, Qinyuan Zheng, Eduardo Ibarra-Garc\'{\i}a-Padilla, Richard~T.
  Scalettar, and Kaden R.~A. Hazzard.
\newblock Unit-density su(3) fermi-hubbard model with spin-flavor imbalance.
\newblock {\em Phys. Rev. A}, 112:033313, Sep 2025.

\bibitem{Pearce_1975}
Paul~A. Pearce and Colin~J. Thompson.
\newblock The anisotropic heisenberg model in the long-range interaction limit.
\newblock {\em Communications in Mathematical Physics}, 41(2):191--201, 1975.

\bibitem{Blin_1996}
Alex~H Blin, Brigitte Hiller, and Li~Junqing.
\newblock Tunnelling at finite temperature in the lmg model.
\newblock {\em Journal of Physics A: Mathematical and General}, 29(14):3993,
  jul 1996.

\bibitem{Das_2006}
Arnab Das, K.~Sengupta, Diptiman Sen, and Bikas~K. Chakrabarti.
\newblock Infinite-range ising ferromagnet in a time-dependent transverse
  magnetic field: Quench and ac dynamics near the quantum critical point.
\newblock {\em Phys. Rev. B}, 74:144423, Oct 2006.

\bibitem{Wilms_2012}
Johannes Wilms, Julien Vidal, Frank Verstraete, and SÃ©bastien Dusuel.
\newblock Finite-temperature mutual information in a simple phase transition.
\newblock {\em Journal of Statistical Mechanics: Theory and Experiment},
  2012(01):P01023, jan 2012.

\bibitem{Dusuel_2005}
S\'ebastien Dusuel and Julien Vidal.
\newblock Continuous unitary transformations and finite-size scaling exponents
  in the lipkin-meshkov-glick model.
\newblock {\em Phys. Rev. B}, 71:224420, Jun 2005.

\bibitem{Campbell_2015}
Steve Campbell, Gabriele De~Chiara, Mauro Paternostro, G.~Massimo Palma, and
  Rosario Fazio.
\newblock Shortcut to adiabaticity in the lipkin-meshkov-glick model.
\newblock {\em Phys. Rev. Lett.}, 114:177206, May 2015.

\bibitem{Pal_2023}
Kunal Pal, Kuntal Pal, and Tapobrata Sarkar.
\newblock Complexity in the lipkin-meshkov-glick model.
\newblock {\em Phys. Rev. E}, 107:044130, Apr 2023.

\bibitem{Pagano_2015}
G.~Pagano, M.~Mancini, G.~Cappellini, L.~Livi, C.~Sias, J.~Catani, M.~Inguscio,
  and L.~Fallani.
\newblock Strongly interacting gas of two-electron fermions at an orbital
  feshbach resonance.
\newblock {\em Phys. Rev. Lett.}, 115:265301, Dec 2015.

\bibitem{Hofer_2015}
M.~H\"ofer, L.~Riegger, F.~Scazza, C.~Hofrichter, D.~R. Fernandes, M.~M.
  Parish, J.~Levinsen, I.~Bloch, and S.~F\"olling.
\newblock Observation of an orbital interaction-induced feshbach resonance in
  $^{173}\mathrm{Yb}$.
\newblock {\em Phys. Rev. Lett.}, 115:265302, Dec 2015.

\bibitem{Rapp_2007}
\'Akos Rapp, Gergely Zar\'and, Carsten Honerkamp, and Walter Hofstetter.
\newblock Color superfluidity and ``baryon'' formation in ultracold fermions.
\newblock {\em Phys. Rev. Lett.}, 98:160405, Apr 2007.

\bibitem{Inaba_2009}
Kensuke Inaba and Sei-ichiro Suga.
\newblock Finite-temperature properties of attractive three-component fermionic
  atoms in optical lattices.
\newblock {\em Phys. Rev. A}, 80:041602, Oct 2009.

\bibitem{Titvinidze_2011}
I~Titvinidze, A~Privitera, S-Y Chang, S~Diehl, M~A Baranov, A~Daley, and
  W~Hofstetter.
\newblock Magnetism and domain formation in su(3)-symmetric multi-species fermi
  mixtures.
\newblock {\em New Journal of Physics}, 13(3):035013, mar 2011.

\bibitem{Pohlmann_2013}
J.~Pohlmann, A.~Privitera, I.~Titvinidze, and W.~Hofstetter.
\newblock Trion and dimer formation in three-color fermions.
\newblock {\em Phys. Rev. A}, 87:023617, Feb 2013.

\bibitem{Chetcuti_2023}
Wayne~Jordan Chetcuti, Andreas Osterloh, Luigi Amico, and Juan Polo.
\newblock {Interference dynamics of matter-waves of SU($N$) fermions}.
\newblock {\em SciPost Phys.}, 15:181, 2023.

\bibitem{Chetcuti_2023b}
Wayne~J. Chetcuti, Juan Polo, Andreas Osterloh, Paolo Castorina, and Luigi
  Amico.
\newblock Probe for bound states of su(3) fermions and colour deconfinement.
\newblock {\em Communications Physics}, 6(1):128, 2023.

\bibitem{Xu_2023}
Han Xu, Xiang Li, Zhichao Zhou, Xin Wang, Lei Wang, Congjun Wu, and Yu~Wang.
\newblock Trion states and quantum criticality of attractive su(3) dirac
  fermions.
\newblock {\em Phys. Rev. Res.}, 5:023180, Jun 2023.

\bibitem{Li_2025}
Xiang Li, Yumeng Li, Quan Fu, and Yu~Wang.
\newblock Trion ordering in the attractive three-color hubbard model on a
  $\ensuremath{\pi}$-flux square lattice.
\newblock {\em Phys. Rev. A}, 112:063319, Dec 2025.

\bibitem{Stepp_2025}
Jonathan Stepp, Eduardo Ibarra-Garcia-Padilla, Richard~T. Scalettar, and Kaden
  R.~A. Hazzard.
\newblock Trion formation and ordering in the attractive su(3) fermi-hubbard
  model.
\newblock 2025.

\bibitem{hazzard}
Kaden R.~A. Hazzard, Victor Gurarie, Michael Hermele, and Ana~Maria Rey.
\newblock High-temperature properties of fermionic alkaline-earth-metal atoms
  in optical lattices.
\newblock {\em Phys. Rev. A}, 85:041604, Apr 2012.

\bibitem{Yanatori_2016}
Hiromasa Yanatori and Akihisa Koga.
\newblock Finite-temperature phase transitions in the $\mathrm{SU}(n)$ hubbard
  model.
\newblock {\em Phys. Rev. B}, 94:041110, Jul 2016.

\bibitem{He_2025}
Chengdong He, Xin-Yuan Gao, Ka~Kwan Pak, Yu-Jun Liu, Peng Ren, Mengbo Guo,
  Entong Zhao, Yangqian Yan, and Gyu-Boong Jo.
\newblock Thermodynamics of spin-imbalanced fermi gases with
  $\mathrm{S}\mathrm{U}(n)$-symmetric interaction.
\newblock {\em Phys. Rev. Lett.}, 134:183406, May 2025.

\bibitem{Capponi_2025}
Sylvain Capponi, Lukas Devos, Philippe Lecheminant, Keisuke Totsuka, and
  Laurens Vanderstraeten.
\newblock Non-landau quantum phase transition in modulated su($n$) heisenberg
  spin chains.
\newblock {\em Phys. Rev. B}, 111:L020404, Jan 2025.

\bibitem{Kobayashi_2012}
Keita Kobayashi, Masahiko Okumura, Yukihiro Ota, Susumu Yamada, and Masahiko
  Machida.
\newblock Nontrivial haldane phase of an atomic two-component fermi gas trapped
  in a 1d optical lattice.
\newblock {\em Phys. Rev. Lett.}, 109:235302, Dec 2012.

\bibitem{Bois_2015}
V.~Bois, S.~Capponi, P.~Lecheminant, M.~Moliner, and K.~Totsuka.
\newblock Phase diagrams of one-dimensional half-filled two-orbital
  $\mathrm{SU}(n)$ cold fermion systems.
\newblock {\em Phys. Rev. B}, 91:075121, Feb 2015.

\end{thebibliography}

\appendix

\section{Harmonic approximation in the broken symmetry phase}\label{app:HP}
To get the low-energy spectrum in the broken symmetry phase, we need to derive the second-order contribution in the Holstein-Primakoff expansion.
For that, we first expand 
\begin{multline}
 \sqrt{N-\sum_{j<L} a_j^{\dag} a_j} \approx \sqrt{N} \mu_L \left(1 - \frac{1}{2\sqrt{N}\mu_L^2} \sum\limits_{j} \mu_j \delta_j^\dagger + \mu_j^* \delta_j \right. \\
 \left. - \frac{1}{2N\mu_L^2} \left(\sum\limits_{j} \delta^\dag_j \delta_j + \sum\limits_{j, k} \frac{( \mu_j \delta_j^\dagger + \mu_j^* \delta_j)( \mu_k \delta_k^\dagger + \mu_k^* \delta_k) }{4 \mu_L^2} \right) \right)   
\end{multline}
To simplify notations, we note:
\begin{multline}
 \sqrt{N-\sum_{j<L} a_j^{\dag} a_j} \approx \sqrt{N} (\mu_L -\frac{1}{\sqrt{N}}A - \frac{B}{N}) ,
\end{multline}

The kinetic term simply becomes
\begin{align}
    H_K&=2\mu_L^2(2B-A^2) - 2\sum\limits_k \cos \frac{2k\pi}{L} \delta^\dagger_k \delta_k \nonumber \\
    &=  2\sum\limits_k (1-\cos \frac{2k\pi}{L}) \delta^\dagger_k \delta_k
\end{align}
For the interaction terms, we obtain
\begin{multline}
   \frac{L}{N}  H_I = -2B\sum\limits_{k, q} (\mu_k \mu_q^*\mu_{k-q}^* + \mu_{k}\mu_{q}\mu_{k+q}^*) \\
   + A^2 \sum\limits_{q} (\mu_q\mu_{-q} + \mu_q^*\mu_{-q}^*)\\
   -2  A \sum\limits_{q, k\neq 0} (\delta_k \mu^*_q \mu^*_{k-q} + \delta^\dagger_k \mu_q \mu_{k-q})\\
   +\sum\limits_{k\neq 0, k'\neq 0, q} (\delta_k \delta_{k'} \mu_{q}^* \mu_{k+k'-q}^* + \delta_k^\dagger \delta^\dagger_{k'} \mu_{q} \mu_{k+k'-q}) \\
   +4 A^2 \sum\limits_{k} \vert \mu_k\vert^2 + 4 \sum\limits_{k\neq0, k'\neq 0, q} \delta^\dagger_{k} \delta_{k'} \mu_{q} \mu_{k'+q-k}\\
   -4A \sum\limits_{k\neq 0, q} \mu_q \delta^\dagger_k \mu^*_{q-k} + \mu_q^* \delta_k \mu_{q-k}
\end{multline}
We can rewrite the Hamiltonian as
\begin{equation}
    \frac{L}{N}H_I =  \sum\limits_{i, j} M_{i, j} \delta^\dagger_i \delta_j + \Delta_{i, j} \delta_i \delta_j + \Delta_{j, i}^* \delta^\dagger_i \delta^\dagger_j,
\end{equation}
\begin{multline}
   \frac{M_{i, j}}{ \mu^*_j \mu_i } = \frac{\Delta_2}{2\mu_L^2} -\frac{M_3}{2\mu_L^3} - \delta_{i, j}\frac{M_3}{\mu_L \vert \mu_j\vert^2} -\sum\limits_{k} \frac{\mu_k^*\mu_{j-k}^*}{\mu_j^* \mu_L}\\
   -\sum\limits_{k} \frac{\mu_k\mu_{i-k}}{\mu_i \mu_L} + \frac{2}{\mu_L^2}-2\sum\limits_{k}\frac{\mu^*_k \mu_{k-j}}{\mu_L\mu_j^*}-2\sum\limits_{k}\frac{\mu_k \mu^*_{k-i}}{\mu_L\mu_i}\\
   +4\sum\limits_k \frac{\mu^*_{k+j-i} \mu_k}{\mu^*_j \mu_i}
\end{multline}
\begin{multline}
   \frac{\Delta_{i, j}}{ \mu^*_j \mu^*_i } = \frac{\Delta_2}{4\mu_L^2} -\frac{M_3}{4\mu_L^3} -\frac{1}{2}\sum\limits_{k} (\frac{\mu_k^*\mu_{j-k}^*}{\mu_j^* \mu_L} + \frac{\mu_k^*\mu_{i-k}^*}{\mu_i^* \mu_L})\\
   + \frac{1}{\mu_L^2}-\sum\limits_{k}(\frac{\mu^*_k \mu_{k-j}}{\mu_L\mu_j^*}+\frac{\mu^*_k \mu_{k-i}}{\mu_L\mu_i^*}) +\sum\limits_k \frac{\mu^*_{i+j-k} \mu^*_k}{\mu^*_j \mu^*_i}
\end{multline}
where we denoted
\begin{equation}
    M_3 = \sum\limits_{k, q} \mu_k \mu_q^*  \mu_{k-q}^* + \mu^*_k \mu_q \mu_{k-q}
\end{equation}
\begin{equation}
    \Delta_2 = \sum\limits_{k} \mu_k \mu_{-k} + \mu_k^* \mu_{-k}^*
\end{equation}
Now, we can numerically prove that we can take both $\mu$ real and assume $\mu_k = \mu_{-k}$ at the minimal configuration (as numerically obtained).
That means that 
\begin{equation}
    \Delta_2 = 2, \quad  M_3 = 2\sum\limits_{k, q} \mu_k \mu_q  \mu_{k-q}
\end{equation}
\begin{multline}
   \frac{M_{i, j}}{ \mu_j \mu_i } = \frac{3}{\mu_L^2} -\frac{M_3}{2\mu_L^3} - \delta_{i, j}\frac{M_3}{\mu_L \vert \mu_j\vert^2} \\ -3\sum\limits_{k} (\frac{\mu_k\mu_{j-k}}{\mu_j \mu_L}+ \frac{\mu_k\mu_{i-k}}{\mu_i \mu_L}) +4\sum\limits_k \frac{\mu_{k+j-i} \mu_k}{\mu_j \mu_i}
\end{multline}
\begin{multline}
   \frac{\Delta_{i, j}}{ \mu_j \mu_i } = \frac{3}{2\mu_L^2} -\frac{M_3}{4\mu_L^3} -\frac{3}{2}\sum\limits_{k} (\frac{\mu_k\mu_{j-k}}{\mu_j \mu_L} + \frac{\mu_k\mu_{i-k}}{\mu_i \mu_L})\\
   +\sum\limits_k \frac{\mu_{i+j-k} \mu_k}{\mu_j \mu_i}
\end{multline}
Solving the one-body Hamiltonian through a Bogoliubov transform gives us the low-energy spectrum of the model, with $\{\mu_k = \mu_k^m\}$ the energy-minimizing configuration.\\

\section{Hartree-Fock development}\label{app:MF}
In this section, we show that the mean-field computation agree with the results obtained from the HP transformation.
We start from the Hamiltonian in Eq.~\eqref{eq:Ham}.
Using $n_{j, \sigma}^2 = n_{j, \sigma}$, we rewrite it as
\begin{equation}
    H = -t \sum\limits_{i} \left(E_{i, i+1} + h.c. \right) + \frac{U}{2} \sum\limits_{i}\left( E_{i, i} + \sum\limits_{\sigma \neq \sigma'} n_{i, \sigma} n_{i, \sigma'} \right).
\end{equation}
We then use the mean-field decoupling
\begin{equation}
    n_{j, \sigma}n_{j, \sigma'} \approx   n_{j, \sigma} \langle n_{j, \sigma'} \rangle +    \langle n_{j, \sigma}  \rangle n_{j, \sigma'} -    \langle n_{j, \sigma}  \rangle  \langle n_{j, \sigma'} \rangle,
\end{equation}
to rewrite the interaction as 
\begin{align}
    & \sum\limits_{i}\left( E_{i, i} + \sum\limits_{\sigma \neq \sigma'} \langle n_{i, \sigma} \rangle n_{i, \sigma'} + \langle n_{i, \sigma'} \rangle n_{i, \sigma} - \langle n_{i, \sigma} \rangle \langle  n_{i, \sigma'} \rangle \right) \nonumber \\
    &= \sum\limits_{i}E_{i, i} + 2 \sum\limits_{\sigma}(\langle E_{i, i}\rangle - \langle n_{i, \sigma}\rangle )n_{i, \sigma}  \nonumber \\ &-\sum\limits_{\sigma}(\langle E_{i, i}\rangle - \langle n_{i, \sigma}\rangle)\langle n_{i, \sigma} \rangle.
\end{align}
This mean-field  respects the U($N$) symmetry of the model.
Now we perform a first approximation: we assume that $\langle n_{i, \sigma} \rangle$ is independent of $\sigma$, which we will note $\langle n_i\rangle$.
This implies that the groundstate is unbroken, i.e. has a singlet magnetic component.
%
%
We simplify the interaction into:
\begin{equation}
     \sum\limits_{i}E_{i, i} + 2 (N - 1) \langle n_{i} \rangle E_{i, i} - N (N-1) \langle n_{i}\rangle \langle n_{i} \rangle.
\end{equation}
We can note that there is full decoupling of the different spin flavour.
It is enough therefore to self-consistently solve the simple $L$-site Hamiltonian
\begin{align}
H &= -t \sum\limits_{i} \left(c^\dagger_{i} c_{i+1} + h.c. \right) + \frac{U}{2}  \nonumber \\
&+U (N-1) \sum\limits \langle n_{i} \rangle ( n_{i} - \frac{1}{2}\langle n_{i} \rangle ).
\end{align}
We included constants for completeness.
It is convenient to define $\langle \delta n_{j} \rangle = \langle n_{j} \rangle - \frac{1}{L}$, such that 
\begin{multline}
H \equiv -t \sum\limits_{i} \left(c^\dagger_{i} c_{i+1} + h.c. \right)  + U (N-1) \sum\limits \langle \delta n_{i} \rangle  n_{i}  \\- \frac{U (N-1)}{2} \sum\limits \langle n_{i} \rangle^2.
\end{multline}

\begin{itemize}
    \item $L = 2$. We use $\langle \delta n_{1} \rangle  = - \langle \delta  n_{2} \rangle $, so that, we only need to solve 
\begin{equation}
    \begin{pmatrix}
        U (N-1)\langle \delta  n_1\rangle &  -t\\
        -t & - U (N-1)\langle \delta  n_1\rangle
    \end{pmatrix}
\end{equation}
whose eigenenergies are 
\begin{equation}
    E_\pm = \pm \sqrt{U^2 (N-1)^2\langle \delta n_1\rangle^2 + t^2}.
\end{equation}
We define 
\begin{equation}
    \begin{cases}
        \cos \theta  = \frac{U (N-1)\langle \delta n_1\rangle}{E_+} \\
        \sin \theta  = \frac{t}{E_+} 
    \end{cases}
\end{equation}
The groundstate is given by $(-\sin \theta/2, \cos \theta/2)$, and therefore, 
\begin{equation}
\langle n_1 \rangle = \sin^2 \theta/2  = \frac{1}{2} - \frac{U (N-1)\langle \delta  n_1\rangle}{2E_+} 
\end{equation}

\begin{equation}
\langle \delta n_1 \rangle = - \frac{U (N-1)\langle \delta  n_1\rangle}{2E_+} 
\end{equation}
$\langle \delta n_1 \rangle = 0$ is always solution of the equation, as expected from the infinite temperature solution.
The other solution (only for $U < 0$) satisfies
\begin{align}
4 E_+^2 &= U^2 (N-1)^2 \\ U^2 (N-1)^2\langle \delta n_1\rangle^2 + t^2 &= \frac{U^2 (N-1)^2}{4}
\end{align}
This solution only exists when $U \leq U_C = -\frac{2}{N-1}$, and
\begin{equation}
\langle \delta n_1\rangle_\pm = \pm \frac{1}{2} \sqrt{1 - \frac{4t^2}{U^2(N-1)^2}}
\end{equation}
These two solutions are equivalent, exchanging the two sites.
Finally, we can compare energies of $ \langle \delta n_1\rangle_\pm$ = $ \langle  \delta n_1\rangle = 0$, and show that we indeed get a continuous meanfield transition for $L = 2$ at the predicted $U_c$.

\item $L = 3$ From inversion symmetry, one can assume $n_2 = n_3$.
We therefore have 
\begin{equation}
    \begin{pmatrix}
        U (N-1)\langle \delta  n_1\rangle &  -t &-t\\
        -t & - \frac{U (N-1)}{2}\langle \delta  n_1\rangle &-t \\
        -t & -t &- \frac{U (N-1)}{2}\langle \delta  n_1\rangle.
    \end{pmatrix}
\end{equation}
$c_- = \frac{1}{\sqrt{2}} (c_2 + c_3)$ decouples from the rest, with energy
\begin{equation}
    E_- = t - \frac{U (N-1)}{2}\langle \delta  n_1\rangle
\end{equation}
The effective two level system we have to solve is 
\begin{equation}
    \begin{pmatrix}
        U (N-1)\langle \delta  n_1\rangle &  -t \sqrt{2}\\
        -t  \sqrt{2} & -t - \frac{U (N-1)}{2}\langle \delta  n_1\rangle
    \end{pmatrix}
\end{equation}
We proceed as before, introducing $\gamma = U (N-1)\langle \delta   n_1\rangle $ for ease of notations, the spectrum gives
\begin{equation}
    E_\pm = \frac{-t}{2} + \frac{\gamma}{4} \pm \frac{\sqrt{3}}{4} \sqrt{12t^2 + 4 \gamma t + 3 \gamma^2}, 
\end{equation}
\begin{equation}
    \begin{cases}
        \cos \theta = \frac{2t + 3\gamma}{\sqrt{3}\sqrt{12t^2 + 4 \gamma t + 3 \gamma^2}} \\
        \sin \theta = \frac{4\sqrt{2} t}{\sqrt{3}\sqrt{12t^2 + 4 \gamma t + 3 \gamma^2}}
    \end{cases}
\end{equation}
And we finally obtain the self-consistent equation:
\begin{equation}
    \langle n_1 \rangle = \frac{1}{2} - \frac{2t + 3\gamma}{2\sqrt{3}\sqrt{12t^2 + 4 \gamma t + 3 \gamma^2}} 
\end{equation}
\begin{equation}
    \langle \delta n_1 \rangle = \frac{1}{6} - \frac{2t + 3\gamma}{2\sqrt{3}\sqrt{12t^2 + 4 \gamma t + 3 \gamma^2}} \label{eq:MF3L}
\end{equation}
We verify again that $\langle \delta n_1 \rangle = 0$ is always solution, with energies $-\frac{t}{2}$ ($\times 2$) and $t$, which match what we expect when $U = 0$.
An analytical form of the solutions of Eq.~\eqref{eq:MF3L} exists, but it is cumbersome.
We verify that additional solutions for $\langle \delta n_1 \rangle \in [-1/3, 2/3]$ exist for $U (N-1) \leq U_c (N-1) \approx -3.9415$.
Importantly, the disappearance of these solutions correspond to a fusion of two real solutions that split in the complex plane.
The fusion does not occur at $\langle \delta n_1 \rangle = 0$, but $\langle \delta n_1 \rangle \approx 0.252$.
Given that in the limit $U \rightarrow-\infty$, $\langle \delta n_1 \rangle \rightarrow 2/3$, this implies a first-order transition at $T = 0$ within the mean-field approximation.
%
%
Taking into account the energy shift to take the solution that minimizes the global energy, the first transition occurs at marginally smaller $U_c \approx -4.00$.
%

\end{itemize}

\end{document}